\documentclass[%
 reprint,
 amsmath,amssymb,
 aps,
]{revtex4-1}

\pdfoutput=1
\usepackage{graphicx}
\usepackage{epstopdf}
\usepackage{dcolumn}
\usepackage{bm}

\usepackage{subfigure}

\usepackage{amsmath,amssymb,bbm}
\usepackage[usenames,dvipsnames]{xcolor} 

\newcommand{\new}[1]{{\color{Black}} #1}

\DeclareMathOperator*{\E}{\mathbb{E}}
\newcommand{\indicator}[1]{\mathbbm{1}_{#1}}
\newcommand{\ER}{Erd\H{o}s-R\'{e}nyi }
\DeclareMathOperator*{\argmin}{arg\,min}

\newcommand{\jun}{J}
\newcommand{\sen}{S}
\newcommand{\jacobian}{\mathcal{J}}
\newcommand{\phiex}{\hat\phi}
\newcommand{\thetaun}{\bar\theta}
\newcommand{\meanDegreeInEachLayer}{z}

\begin{document}

\preprint{APS/123-QED}

\title{Cascades in multiplex financial networks with debts of different seniority}


\author{Charles D.\ Brummitt}
\affiliation{Center for the Management of Systemic Risk, Columbia University, New York, NY 10027, USA}
\author{Teruyoshi Kobayashi}
\email{Corresponding author. kobayashi@econ.kobe-u.ac.jp}
\affiliation{Graduate School of Economics, Kobe University, 2-1 Rokkodai, Nada, Kobe 657-8501, Japan}

\date{\today}

\begin{abstract}
The seniority of debt, which determines the order in which a bankrupt institution repays its debts, is an important and sometimes contentious feature of financial crises, yet its impact on system-wide stability is not well understood. We capture seniority of debt in a multiplex network, a graph of nodes connected by multiple types of edges. Here, an edge between banks denotes a debt contract of a certain level of seniority. Next we study cascading default. There exist multiple kinds of bankruptcy, indexed by the highest level of seniority at which a bank cannot repay all its debts. Self-interested banks would prefer that all their loans be made at the most senior level. However, mixing debts of different seniority levels makes the system more stable, in that it shrinks the set of network densities for which bankruptcies spread widely. We compute the optimal ratio of senior to junior debts, which we call the \emph{optimal seniority ratio}, for two uncorrelated Erd\H{o}s-R\'{e}nyi networks. 
If institutions erode their buffer against insolvency, then this optimal seniority ratio rises; in other words, if default thresholds fall, then more loans should be senior. We generalize the analytical results to arbitrarily many levels of seniority \new{and to heavy-tailed degree distributions}. 

\begin{description}
\item[PACS numbers]
89.65.Gh, 
89.75.Hc 
\end{description}

\end{abstract}

\maketitle

\section{Introduction}
The global financial crisis of 2007--2009 spurred a flurry of research on financial contagion using a variety of network models~\cite{GaiKapadia2010,Nier2007,Kobayashi2014,Gai2011,Upper2011,May2010,Hurd2013}. In most of these models, financial institutions interact in just one way, such as making one type of loan to one another. In practice, however, banks interact in many ways, all of which can contribute to a crisis: banks issue loans of different durations and of different levels of riskiness; they trade assets with each other; and they hold assets in common. Models have only recently begun to capture more than one of these kinds of interactions~\cite{Bargigli2015,Montagna2013,Elliott2014}. Relatively unexplored is the seniority of debts, which determines the order in which bankrupt institutions repay their debts. 
Individual creditors pay close attention to their assets' seniority levels, but little is known about how the system-wide composition of seniority levels affects the risk of large crises.

A concurrent and mostly independent thread of literature 
in physics has studied multiplex networks, or graphs with multiple sets of edges (or ``network layers'')\new{; for reviews, see}~\cite{Kivela2014_multilayer_review,Boccaletti2014}. 
\new{Variants of percolation on multiplex networks have received much attention~\cite{Buldyrev2010,Son2012,Baxter2012,Cellai2013,Min2014_robustness_multiplex_interlayer_degree_correlations}. 
Closer to financial default contagion are models of threshold cascade on multiplex networks~\cite{Brummitt2012_PRER,Yagan2012,Lee2014}. In these threshold models, nodes exist in one of two states (e.g., adopted a behavior or not, gone bankrupt or not), and after a node adopts the behavior (or goes bankrupt), it influences its neighbors in all layers to do so as well.
}
%
A common conclusion from these models of \new{percolation and threshold cascades on multiplex networks} is that the system's vulnerability to large cascades cannot be understood by examining the vulnerability of any one layer to large cascades\new{~\cite{Buldyrev2010,Son2012,Baxter2012,Brummitt2012_PRER,Yagan2012,Lee2014,Min2014_robustness_multiplex_interlayer_degree_correlations,Cellai2013}}. 
In particular, \new{in threshold models,} the interactions among network layers can make large cascades easier (if nodes combine influence from multiple types of neighbors in a disjunctive~\cite{Brummitt2012_PRER} or weighted~\cite{Yagan2012} way) 
or more difficult 
(if nodes combine influence in a conjunctive way~\cite{Lee2014}).

In this paper, we combine these two strands of research to study a financial phenomena of interest: how  vulnerability to cascading default depends on the seniority of debts. Financial institutions (hereafter called ``banks'' for simplicity) hold debt in one another; more generally, they have exposure to one another's default risk through interbank lending and/or security holding. Each debt contract has a certain level of seniority. When a bank goes into default (i.e., when a bank's liabilities exceed its assets), the bank sells its remaining assets to pay off whatever liabilities it can in decreasing order of seniority, from most senior liability to most junior liability. 

In our model, there exist multiple types of bankruptcy, indexed by the highest level of seniority at which a bank cannot repay all its debts. Using a technique called the cascade condition\new{~\cite{Watts2002,Gleeson2007,Gleeson2008,Payne2011,Yagan2012,Brummitt2012_PRER,Melnik2013,Lee2014}}, we study whether a small number of initial bankruptcies (of certain types) triggers a cascade of many bankruptcies.

The model has interesting tradeoffs. Self-interested banks would prefer to lend at the most senior level in order to maximize the assets they would recover in the event of its debtors defaulting. However, if all banks lend at the most senior level, then the system is especially vulnerable to large cascades of default, in the sense that cascades are likely large for a large interval of network densities. (This extreme case recovers the models in~\cite{GaiKapadia2010,Nier2007}.) 

By contrast, the presence of debts of different seniority makes the system less vulnerable to bankruptcies spreading widely. For the case of two seniority levels (called ``junior'' and ``senior''), our main result is to derive the ``optimal'' ratio of senior loans to junior loans \new{(for a simple class of uncorrelated \ER multiplex networks)}. Here, ``optimal'' means that it minimizes the size of the interval of network densities for which bankruptcies likely spread widely. The optimal multiplex networks have on the order of $50\%$ to $100\%$ more senior loans than junior ones. As banks' leverage (i.e., assets divided by net worth) increases, this ``optimal seniority ratio'' increases, 
meaning that more loans should be senior rather than junior in order to minimize the size of the set of network densities such that cascades are likely large. 
\new{We also show that the optimal seniority ratio is qualitatively similar for heavy-tailed degree distributions like those found in real interbank lending networks~\cite{Boss2004,DeMasi2006,Cont2013}.} 
These results indicate what types of ``seniority--multiplex'' interbank networks are most insusceptible to large cascades of default.

The most recent scheme for financial regulation, called Basel III, emphasizes the liability structure of individual banks, but it does not take into account how seniority of debts affect systemic risk. Our results suggest that seniority of debt could be an important focus of financial regulators and of macro-prudential policies. 

Furthermore, this work contributes to the rapidly growing area of research on contagion in multiplex networks by introducing a model in which nodes can ``activate'' in more than one way, which we call ``multilevel contagion''. 
\new{An interpretation of this model for 
the contagious spread of fads, products, etc.\ in social networks~\cite{Watts2002}  
is that the kinds of neighbors a person influences depends on how excited that person is. Thus, the model resembles the multistage model of complex contagion~\cite{Melnik2013} but on multiplex networks.}

\subsection{Related literature\label{sec:related_literature}}

Theoretical studies of contagion in financial networks generally fall into two classes~\cite{Rogers2013}. Some studies consider a particular instance of a financial network, usually a complicated and in some cases empirically measured object. Other studies consider probability distributions over financial networks.
 
Our study belongs to the latter class, but two papers closely related to our analysis belong to the former~\cite{Elsinger2009, Gourieroux2013}. Elsinger~\cite{Elsinger2009} and Gourieroux et al.~\cite{Gourieroux2013} also assume a senior--junior structure of interbank assets. Whereas they obtain a clearing payment vector for a given network (as in Eisenberg and Noe~\cite{Eisenberg2001}), we study the cascade condition for ensembles of random networks in the same spirit as Watts~\cite{Watts2002} and followup work in physics~\cite{Gleeson2007,Gleeson2008,Centola2007,Payne2009,Payne2011,Liu2012,Melnik2013,Brummitt2012_PRER,Yagan2012,Lee2014}. The former approach can be useful to regulators faced with a real system on the verge of a crisis, whereas the latter approach offers elegant theory and indicative suggestions~\cite{Rogers2013}. 
Although financial networks change over time, there is evidence that their statistical properties (such as degree distributions) can be somewhat stable~\cite{Cont2013}. 

\subsection{Model}

\subsubsection{Multiplex network of loans of different seniority}

We consider a multiplex network consisting of $N$ nodes and $M$ layers. Each node represents a financial institution (or a ``bank'' for simplicity). Each layer is a set of directed edges (i.e., ordered pairs of nodes) that denote loans
and, more generally, credit exposures between banks.  The layer to which an edge belongs is the loan's seniority, and the vector of layers is sorted in increasing order of seniority. When a bank goes bankrupt, it repays its liabilities in decreasing order of seniority; that is, it repays its most senior liabilities first, its second-most senior liabilities second, etc. For the moment, we consider a two-layer (``duplex'') network, and the $M = 2$ many levels of seniority are called ``junior'' and ``senior'' (labeled $\jun$ and $\sen$), respectively. 

The number of banks to which a bank lends funds of seniority $\alpha \in \{\jun , \sen\}$ (i.e., the bank's out-degree in layer $\alpha$) is denoted by $l_\alpha$. Likewise, the number of banks from which a bank borrows funds of seniority $\alpha$ (i.e., the bank's in-degree in layer $\alpha$) is denoted by $b_\alpha$. For simplicity, we assume that the sizes of loans (i.e., the notional values) are the same, so we can consider the networks as unweighted. Thus, the out-degree and the in-degree correspond to the bank's total volume of lending and borrowing, respectively.

\begin{figure}
\includegraphics[width=\columnwidth,clip]{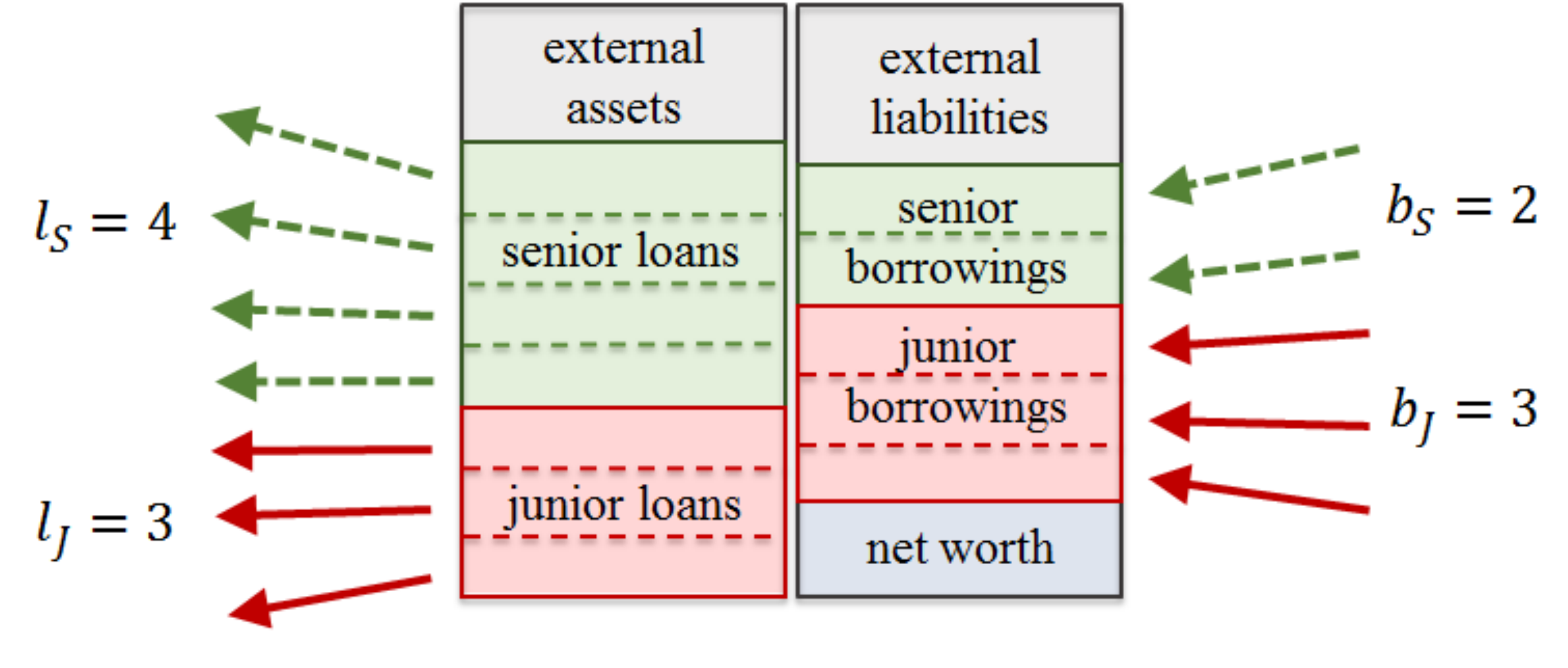}
\caption{(Color online) An example of a bank's balance sheet. Assets (left-hand column) consist of external assets, $l_\sen = 4$ many senior interbank loans, and $l_\jun = 3$ many junior interbank loans. Liabilities (right-hand column) consist of external liabilities, $b_\sen = 2$ many senior interbank liabilities, and $b_\jun = 3$ many junior interbank liabilities. Net worth (a.k.a.\ equity or, for banks, capital) is the difference between assets and liabilities.}
\label{fig:balance_sheet}
\end{figure}

Banks may also hold  assets and issue liabilities outside the banking system. (For instance, external creditors may lend to banks in the form of demand deposits.) The balance sheet of a typical bank is illustrated in Fig.~\ref{fig:balance_sheet}.

\subsubsection{Multiple types of default}

In a model with two levels of seniority of debt, there are two types of defaults (bankruptcies): junior-defaults and senior-defaults. A bank is in \emph{junior-default} if and only if its capital (a.k.a.\ equity or net worth; see Fig.~\ref{fig:balance_sheet}) is negative, meaning that it cannot pay all of its junior debts. A bank is in \emph{senior-default} if and only if its capital is so far below zero that it cannot pay its all of its senior debts nor any of its junior debts. Note that banks in senior-default are necessarily in junior-default as well. 

We assume that when a bank goes into junior-default, it cannot repay any of its junior-level liabilities, and when a bank goes into senior-default, it cannot pay any of its junior liabilities nor any of its senior liabilities. Said differently, we assume that loss given default of seniority $\alpha$ is $100\%$ for interbank liabilities of seniority of $\alpha$ and lower. This assumption departs from previous cascade models of financial contagion, such as that of Gai and Kapadia~\cite{GaiKapadia2010}, in which loss given default is $100\%$ for all interbank liabilities, no matter their seniority. Table~\ref{tab:value_of_debt} makes this assumption concrete: the value of a bank's debt is equal to some positive value (say, $1$ unit of currency) if the bank is solvent (i.e., if its equity is nonnegative); the value of the bank's junior debt falls to zero when the bank becomes insolvent, and the value of its senior debt falls to zero when the bank's equity passes a nonpositive threshold value (derived later to be $-b_\jun$).

\begin{table}[htb]
\caption{Relationship between the equity of a borrower and the value of its junior debts and senior debts held by its creditors. Here, we assume that junior and senior debts are equally valuable when the borrower is solvent, so values of debts are either zero or $\$1$. If the borrower's equity is negative but sufficiently large ($\geq -b_\jun$), then it can pay its senior debts but none of its junior debts; the borrower is in ``junior-default''. For sufficiently negative equity of the borrower, the borrower cannot pay neither its junior debts nor its senior debts; the borrower is in ``senior-default'', and all its debts have value zero.}
\begin{center}
\begin{tabular}{c|c|c|c}
equity & status & junior debt value & senior debt value \\
 \hline
$\geq 0$ & solvent & \$1 & \$1 \\
$< 0$& junior-default & 0 & \$1 \\
$< -b_\jun$ & senior-default & 0 & 0
\end{tabular}
\end{center}
\label{tab:value_of_debt}
\end{table}%

\subsubsection{Thresholds for junior- and senior-default}
\label{sec:cascades_description}

Initially, all banks are solvent, and a small fraction of banks are chosen to be in junior-default or in senior-default. These initial defaults could be caused by the loss of external assets. These bank failures cause losses for their creditors, who may default as a result, which may cause one or more of their creditors to default, and so on, resulting in a cascade of default.

Consider some point in time during a cascade. For a certain bank, let $m_\jun$ denote the number of its junior-borrowers that have junior-defaulted, and let $m_\sen$ denote the number of its senior-borrowers that have senior-defaulted. This bank is in junior-default if and only if its losses $m_\jun + m_\sen$ exceeds the bank's equity $w$, and this bank is in senior-default if and only if its losses $m_\jun + m_\sen$ exceed the sum of its equity $w$ and its junior liabilities $b_\jun$. 

\begin{figure*}
 \begin{center}
\includegraphics[width=1.6\columnwidth,clip]{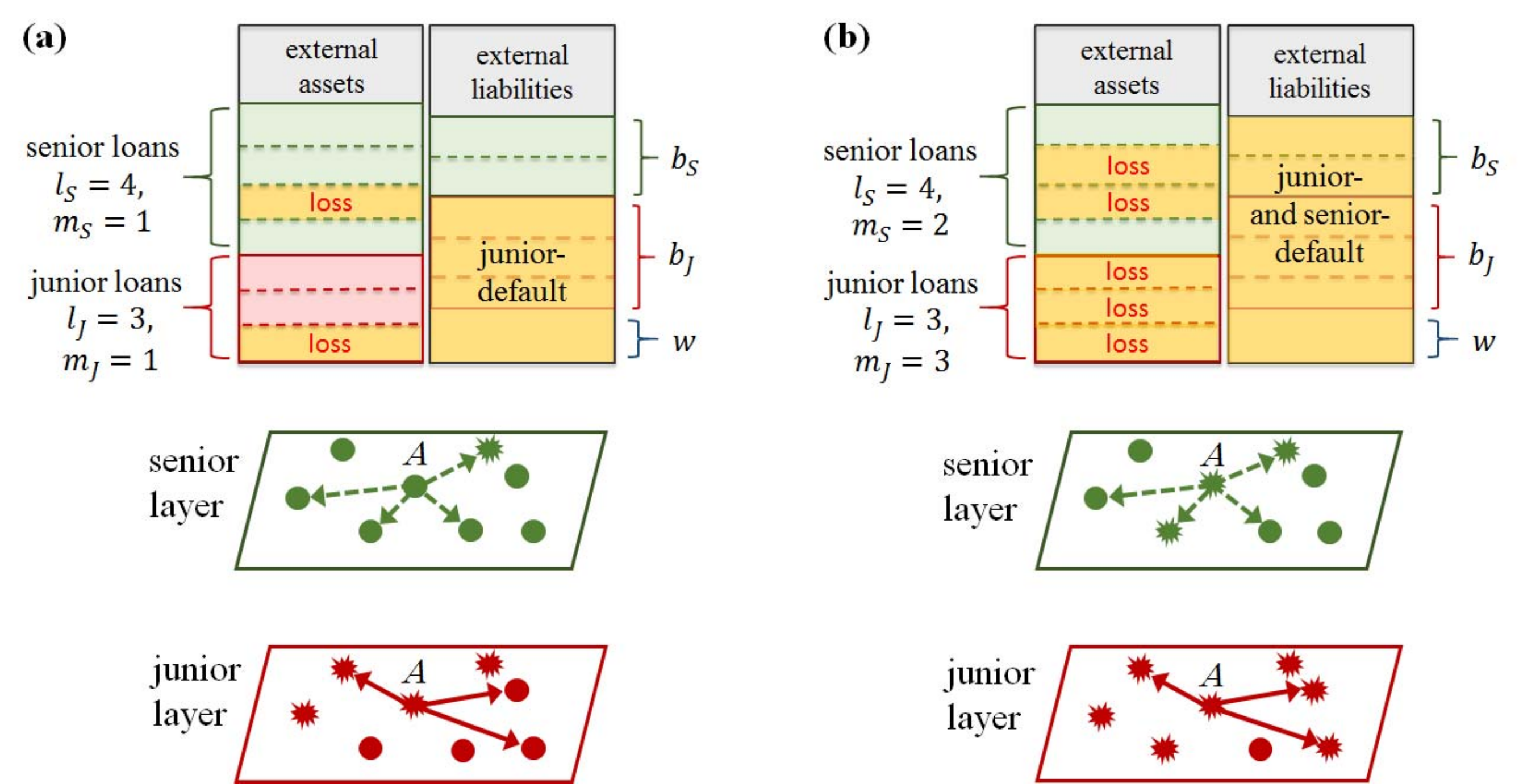} 
  ~\caption{(Color online) An example of the balance sheet  (and the corresponding multiplex network visualizations) of a bank that is (a) in junior-default: $m_{\jun}+m_{\sen} >R_\jun (l_{\jun}+l_{\sen}) \equiv {w}$ and (b) in senior-default: $m_{\jun}+m_{\sen} > R_\sen (l_{\jun}+l_{\sen}) \equiv w+b_{\jun}$. Circular and spiky-shaped nodes are solvent and insolvent (a.k.a.\ bankrupt, in default), respectively, in that layer. Edges point from lender to borrower. Senior loans are dashed green arrows; junior loans are solid red arrows. In panel (a), the bank labeled $A$ suffers losses of $m_\jun+m_\sen = 2$ on its assets side (assets are listed on the left-hand side of the bank's balance sheet) because $m_\jun = 1$ of its ($l_\jun=3$ many) junior debtors are in junior-default and because $m_\sen = 1$ of its ($l_\sen=4$ many) senior debtors are in senior-default. These losses exceed bank $A$'s equity ($w$), so bank $A$ cannot pay off all its ($b_\jun$ many) junior liabilities. Because of the assumption that loss given default at layer $\alpha$ is $100\%$ for all creditors in that layer, all of $A$'s junior creditors do not receive repayments on their loan to $A$, and we say that $A$ goes into junior-default. However, bank $A$ does not go into senior-default because it can still pay its senior liabilities (i.e., because $m_\jun+m_\sen \leq w+b_\jun$). In panel (b), bank $A$ suffers so many losses ($m_\jun+m_\sen =5$) that it cannot repay its junior creditors nor its senior creditors, so bank $A$ goes into senior-default.}
  \label{fig:BS_defaults}
 \end{center}
\end{figure*}

To exploit the solution technique for threshold cascades in networks~\cite{Gleeson2007,Gleeson2008,Brummitt2012_PRER,Yagan2012,Lee2014}, we express these inequalities 
in terms of losses relative to total out-degree $l_\jun + l_\sen$: the bank is in default at layer $\alpha \in \{\jun, \sen\}$ if and only if
\begin{align}
 \frac{m_{\jun} + m_{\sen}}{l_{\jun} + l_{\sen}} > R_\alpha \equiv 
 \begin{cases}
 \frac{w}{l_\jun + l_\sen} &\text{if } \alpha = \jun  \\
 \frac{w + b_\jun}{l_\jun + l_\sen} &\text{if } \alpha = \sen
 \end{cases}
 \label{eq:define_thresholds}
\end{align}
[In Appendix~\ref{appendix:thresholds_for_default_many_seniority_levels}, we generalize the thresholds $R_\alpha$ in Eq.~\eqref{eq:define_thresholds} to the general case of $M \geq 1$ layers.] Figure~\ref{fig:BS_defaults} illustrates a bank in junior-default and a bank in senior-default in two ways: as losses on a balance sheet (top row) and as contagion in a multiplex network (bottom row). 

The dynamics of multilevel default contagion can be described as a multiplex version of the threshold model~\cite{Granovetter1978,Watts2002,Centola2007,Payne2011}. (Other multiplex generalizations were studied in~\cite{Brummitt2012_PRER,Yagan2012,Lee2014}.) The ``response functions'' for each layer are 
   \begin{subequations}
   \begin{align} 
   F_{\jun}(\vec l, \vec b, \vec m) 
   &\equiv \begin{cases} 1 & \text{if } \frac{m_{\jun}+m_{\sen}}{l_{\jun}+l_{\sen}} > R_{\jun} \\ 0 & \text{otherwise} \end{cases} \label{eq:define_FJ} \\
   F_{\sen}(\vec l, \vec b, \vec m) 
   &\equiv \begin{cases} 1 & \text{if } \frac{m_{\jun}+m_{\sen} -b_{\jun}}{l_{\jun}+l_{\sen}} > R_{\jun} \\ 0 & \text{otherwise} \end{cases},  \label{eq:define_FS}
 \end{align}
 \label{eq:define_response_functions}
 \end{subequations}
\!\!\!\!\!\! where $\vec l \equiv (l_\jun, l_\sen), \vec b \equiv (b_\jun, b_\sen)$ and $\vec m \equiv (m_\jun, m_\sen)$. If the outputs of $F_\jun$ and $F_\sen$ are both $0$, then the bank is solvent; if $F_\jun$ gives $1$, then the bank is in junior-default; if $F_\sen$ gives $1$, then the bank is in senior-default (and in junior-default).

We note in passing that a special case of this model coincides with a special case of the ``multistage complex contagion" model of Melnik et al.~\cite{Melnik2013}; for details, see Appendix~\ref{sec:relationship_with_multistage}. 
 
\section{Multilevel cascades on 2-layer multiplex networks of junior and senior loans}

This multiplex model adds a new dimension of complexity to previous threshold models~\cite{Watts2002,Gleeson2007,Payne2011,Nier2007,GaiKapadia2010,Kobayashi2014} by considering multiple levels of seniority of debt, so to understand the model we start by simplifying in 
two other dimensions, the distribution of networks and the distribution of thresholds.

First, each layer is modeled as an independent, locally treelike, random directed network generated by the configuration model. The sparsity and independence of the layers implies that the number of ``overlapping edges'' (i.e., junior and senior loans between the same pair of banks) is negligibly small for sufficiently large networks~\cite{Bianconi2013}.

Second, following previous studies~\cite{Nier2007,GaiKapadia2010,Kobayashi2014}, we assume that every bank has the same junior-default threshold $R_{\jun} = w/(l_{\jun}+l_{\sen})$. This assumption implies that all banks have the same ratio of capital to total interbank assets. In the real world, banks' capital-to-asset ratios are not identical, of course, but they are somewhat similar~\cite{BankLeverageData}. It is straightforward to consider heterogeneous capital-to-asset ratios by considering heterogeneous thresholds~\cite{Gleeson2007,Kobayashi2014}.

\subsection{Analytical approximation}
\label{sec:analytical_approximation}

Let $\phi_0^{\jun}$ and $\phi_0^{\sen}$ denote the probabilities that a node is initially in junior-default or in senior-default, where $0 < \phi_0^{\jun}\ll 1$ and $\phi_0^{\sen}\in[0,\phi_0^{\jun}]$. Next we adapt to this model a technique for calculating whether cascades likely spread widely, the \emph{cascade condition}, introduced in~\cite{Watts2002,Gleeson2007} and generalized to other multiplex network models in~\cite{Brummitt2012_PRER,Lee2014,Yagan2012}. 

\subsubsection{Recursion equations}
Consider a bank chosen uniformly at random, and let the network hang down like a tree from this root node by doing a breadth-first search along the directions of the edges. We define $\phi_{t+1}^\alpha$ as the probability that a node $t+1$ hops from the bottom of the tree is in $\alpha$-default because sufficiently many of its debtors (i.e., its out-neighbors) are in default. The nodes at the bottom of the tree are initially in $\alpha$-default with probability $\phi_0^\alpha$. To compute $\phi_{t+1}^\alpha$ from $\phi_t^\alpha$, we condition on whether a node located $t+1$ hops above the leaves of the tree is initially in $\alpha$-default, which occurs independently with probability $\phi_0^\alpha$. If this node is not initially in $\alpha$-default, then we condition on the number of debtors of this node that have defaulted.

For a sufficiently large graph, the fractions of nodes in junior- and senior-default at the end of the cascade are well approximated by a fixed point $(\phi_\infty^\jun, \phi_\infty^\sen)$ of the recursion equations  
\begin{align}
\phi_{t+1}^\alpha &= g^{(\alpha)}(\phi_t^{\jun}, \phi_t^{\sen}) \nonumber \\
&\equiv \phi_0^\alpha + (1-\phi_0^\alpha) \sum_{l_\jun + l_\sen \geq 1} \sum_{b_\jun} p_{l_\jun}^{J,\text{loan}} p_{l_\sen}^{\sen,\text{loan}} p_{b_\jun}^{\jun,\text{borrow}}  
\notag \\ 
& \;\;\; \times  \sum_{m_\jun = 0}^{l_\jun} \sum_{m_\sen = 0}^{l_\sen} B_{m_\jun}^{l_\jun}(\phi_t^\jun) B_{m_\sen}^{l_\sen}(\phi_t^\sen) F_\alpha(\vec l, \vec b, \vec m), \label{eq:recursion_equations}
\end{align}
for $\alpha \in \{\jun, \sen\}$, 
where $p_{b_\alpha}^{\alpha, \text{borrow}}$ and $p_{l_\alpha}^{\alpha, \text{loan}}$ denote the in- and out-degree distributions, respectively, on layer $\alpha \in \{\jun, \sen\}$. In Eq.~\eqref{eq:recursion_equations}, we approximated the numbers of neighbors in junior and senior default as independent 
\new{binomial random variables, the probability mass functions of which are written using the shorthand notation $B_m^l(\phi) \equiv \binom{l}{m} \phi^m (1-\phi)^{l-m}$]. This approximation works well when}
the network layers are independent and hence the number of overlapping edges is negligibly small~\cite{Bianconi2013}. 
\new{(If overlap is significant, then one could adapt the techniques introduced in a study of mutual percolation~\cite{Cellai2013} to this threshold model.)} 
\new{Variants of Eq.~\ref{eq:recursion_equations} appear in several studies of this threshold model; see, for example,\ \cite[Eqs.\ (1)--(3)]{Gleeson2007},\ \cite[Eqs.\ (1) and (4)]{Gleeson2008},\ \cite[Eqs.\ (1)--(3)]{Brummitt2012_PRER},\ \cite[Eqs.\ (11)--(13)]{Yagan2012},\ \cite[Eqs.\ (A9)--(A11)]{Melnik2013},\ \cite[Eqs.\ (1)--(3)]{Lee2014} and, for a rigorous treatment,~\cite[Theorem 1]{Hurd2013}; the closest variant is the study of directed (but not multiplex) networks~\cite[Eqs.\ (14) and (15)]{Payne2011}.}

In short, we approximate the multiplex graph as a tree according to outgoing edges, and  $\phi_{t+1}^\alpha$ is the probability that a bank (located $t+1$ hops from the leaves of the tree) goes into $\alpha$-default \new{because it was initially in $\alpha$-default or} because of the defaults of its debtors. Note from Eq.~\eqref{eq:define_response_functions} that 
\new{going into senior-default due to defaulted debtors implies having gone into junior-default due to defaulted debtors} [i.e., $F_\sen(\vec l, \vec b, \vec m) = 1$ implies $F_\jun(\vec l, \vec b, \vec m) = 1$], 
so \new{the fraction in senior-default never exceeds the fraction in junior-default [i.e.,} $\phi_t^\sen \leq \phi_t^\jun$ for all $t \geq 0$].

For convenience, we rewrite the recursions~\eqref{eq:recursion_equations} as
\begin{align}
\begin{pmatrix}\phi_{t+1}^{\jun} \\ \phi_{t+1}^{\sen} \end{pmatrix} = \begin{pmatrix} g^{(\jun)}(\phi_t^{\jun}, \phi_t^{\sen}) \\ g^{(\sen)}(\phi_t^{\jun}, \phi_t^{\sen}) \end{pmatrix}. \label{eq:recursion_vector}
\end{align}
Iterating Eq.~\eqref{eq:recursion_vector} to a fixed point $(\phi_\infty^\jun, \phi_\infty^\sen)$ gives 
the expected fractions of banks in junior-default and senior-default, respectively, at the end of the cascade. 

\subsubsection{Multiplex cascade condition}
\label{sec:multiplex_cascade_condition}
Now we obtain a simple expression that approximately captures, in the limit of network size approaching infinity, for which parameters a vanishingly small seed triggers a cascade that results in a finite fraction of banks in junior- or senior-default. The \emph{first-order cascade condition} is the linear instability of Eq.~\eqref{eq:recursion_vector} at the origin $(0,0)$, which provides a sufficient condition for the numbers of junior- and senior-defaulted banks to grow. The Jacobian matrix of the right-hand side of Eq.~\eqref{eq:recursion_vector} is denoted by 
\begin{align}
\jacobian \equiv \begin{pmatrix} \jacobian_{\jun\jun} & \jacobian_{\jun\sen} \\ \jacobian_{\sen\jun} & \jacobian_{\sen\sen} \end{pmatrix}. 
 \label{eq:J_entries} 
\end{align}
The first-order cascade condition is that the largest eigenvalue of the Jacobian matrix, $\lambda_{\max}(\jacobian)$, exceeds $1$:
\begin{align}
\lambda_{\max}(\jacobian) &\equiv \frac{\jacobian_{\jun\jun} + \jacobian_{\sen\sen} + \sqrt{(\jacobian_{\jun\jun} - \jacobian_{\sen\sen})^2 + 4 \jacobian_{\jun\sen} \jacobian_{\sen\jun}}}{2} \notag \\ &> 1. \label{eq:FOcascadecondition}
\end{align}
In Appendix~\ref{sec:derive_FOCC}, we simplify the Jacobian matrix and show that, for this model, its largest eigenvalue equals its trace.

\subsection{Cascade regions}
\label{sec:cascade_regions}
\new{
If many banks are in junior-default, the fraction of banks in senior-default could be large or small. Thus, there are two other cascade conditions: one for junior-defaults and another for senior-defaults. 
Next we derive these conditions 
in some generality (for ensembles of multiplex networks with independent layers that are specified by their degree distributions). 
We also illustrate 
all three cascade conditions (junior, senior, and multiplex) 
for a particular case of interest: a multiplex network of two uncorrelated \ER networks of mean degree $\langle l_\jun \rangle$ and $\langle l_\sen \rangle$, respectively.}


\subsubsection{Cascades of junior-default (ignoring senior-default)}
First consider contagion only on junior loans. The ``junior-only'' cascade condition is that a junior-default causes on average at least one more junior-default, or equivalently that the top-left entry of the Jacobian matrix, $\jacobian_{\jun \jun}$, exceeds $1$. In Appendix~\ref{sec:derive_FOCC}, we show that this junior-only cascade condition is the inequality 
\begin{align}
\jacobian_{\jun \jun} &= \E [ l_\jun (\indicator{1 > R_\jun (l_\jun + l_\sen)})] > 1. \label{eq:junior_only_cascade_condition}
\end{align}

The set of parameters satisfying this junior-only cascade condition~\eqref{eq:junior_only_cascade_condition} is shown in orange in Fig.~\ref{fig:cascaderegion}(a) \new{for uncorrelated \ER networks of mean degree $\langle l_\jun \rangle$ and $\langle l_\sen \rangle$}. If the senior-layer is the empty graph (i.e., if $\langle l_\sen \rangle = 0$), then we recover standard cascade models on a single-layer network~\cite{Watts2002,Gleeson2007,Payne2011,Nier2007,GaiKapadia2010,Kobayashi2014}. Creating more senior loans (i.e., increasing $\langle l_\sen \rangle$) shrinks the ``junior-only'' cascade region. This shrinking occurs because a bank's equity $w = R_\jun (l_\jun + l_\sen)$ increases with its number of senior loans $l_\sen$ [recall Eq.~\eqref{eq:define_thresholds}]. Thus, as the expected number $\langle l_\sen \rangle$ of senior loans per bank increases, a typical bank's buffer against junior-default ($w$) increases, which makes a junior-default less likely to cause other junior-defaults.

\subsubsection{Cascades of senior-default (ignoring junior-default)}
In a similar way, consider contagion along senior loans only. The ``senior-only'' cascade condition is that a senior-default causes on average at least one more senior-default, or equivalently that the bottom-right entry of the Jacobian matrix, $\jacobian_{\sen \sen}$, exceeds $1$. In Appendix~\ref{sec:derive_FOCC}, we show that this senior-only cascade condition is 
\begin{align}
\jacobian_{\sen \sen} &= \E [ l_\sen (\indicator{1 - b_\jun > R_\jun (l_\jun + l_\sen)})] \notag \\
&= p_{0}^{\jun, \text{borrow}} \E[l_\sen \indicator{1 > R_\jun (l_\jun + l_\sen)}] 
> 1. \label{eq:senior_only_cascade_condition}
\end{align}
The first-order cascade condition considers defaults triggered by the default of just one out-neighbor (i.e., defaults due to $m_\jun + m_\sen =1$); thus, it captures senior-defaults of banks that lack junior debts ($b_\jun = 0$), which explains the factor $p_{0}^{\jun, \text{borrow}}$ in Eq.~\eqref{eq:senior_only_cascade_condition}. We found that capturing defaults caused by two bankrupt creditors (using the second-order cascade condition) was not necessary to approximate simulations well, as shown in  Appendix~\ref{sec:cascade_region_fixed_point} \new{and in Fig.~\ref{fig:staticmodel}}.

The parameters satisfying this senior-only cascade condition~\eqref{eq:senior_only_cascade_condition} are shown in green in Fig.~\ref{fig:cascaderegion}(a) \new{for uncorrelated \ER networks of mean degree $\langle l_\jun \rangle$ and $\langle l_\sen \rangle$}. This region shrinks with the density of junior loans $\langle l_\jun \rangle$ for the same reason that the junior-only cascade region shrinks with $\langle l_\sen \rangle$. 

\subsubsection{Multiplex cascades}

\begin{figure}[htb]
\begin{center}
\includegraphics{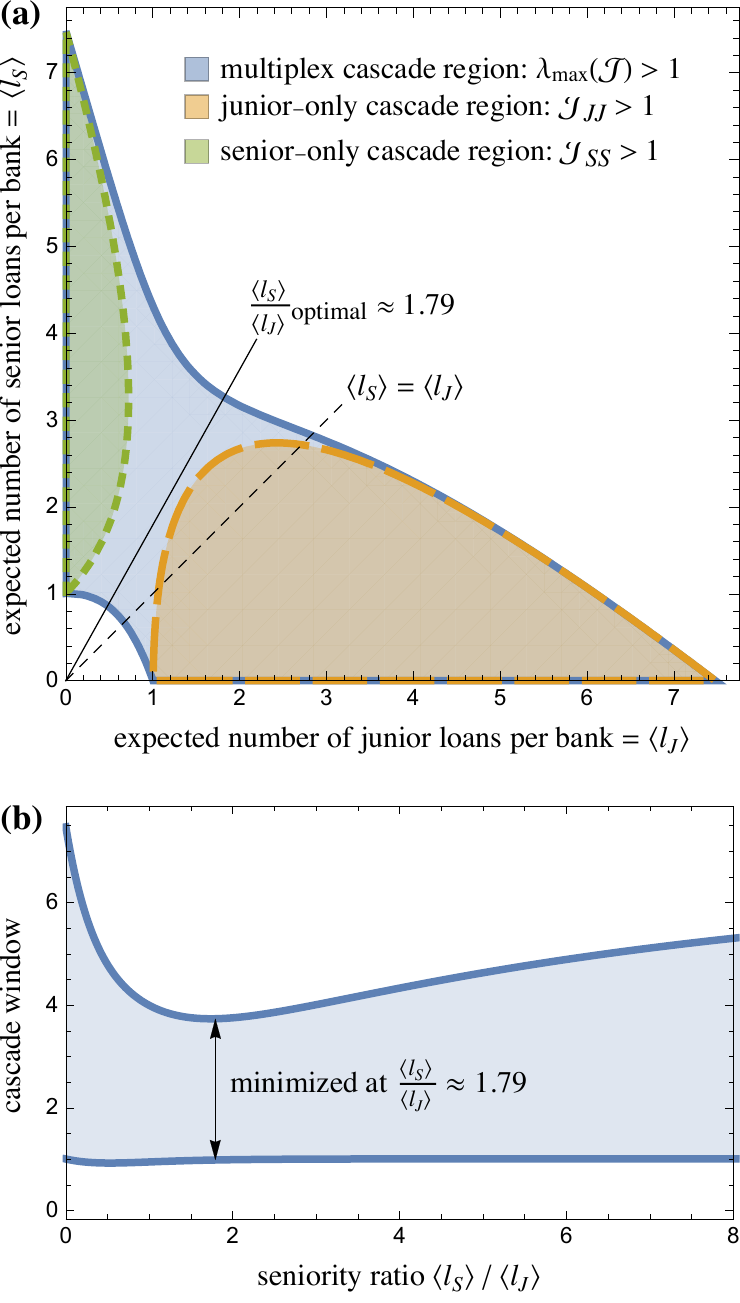}
\caption{
(Color online) 
(a) The multiplex cascade region [the edge-density parameters $(\langle l_\jun \rangle, \langle l_\sen \rangle)$ satisfying Eq.~\eqref{eq:FOcascadecondition}, drawn as a blue region with a thick, solid boundary] contains the junior-only and senior-only cascade conditions given by Eqs.~\eqref{eq:junior_only_cascade_condition} and~\eqref{eq:senior_only_cascade_condition} and drawn as an orange region (with a dashed boundary) and green region (with a dotted boundary), respectively. Moreover, the multiplex cascade region is asymmetric in the densities of junior and senior loans, $(\langle l_\jun \rangle, \langle l_\sen \rangle)$. The optimal seniority ratio, defined in Eq.~\eqref{eq:define_optimal_ratio}, is drawn as a black line. This ratio $\langle l_\sen \rangle / \langle l_\jun \rangle \approx 1.79$ minimizes the cascade window, the set of mean degrees satisfying the first-order cascade condition, Eq.~\eqref{eq:FOcascadecondition}. The case of equally many junior and senior loans (on average) is shown with a dashed line, which has a slightly larger cascade window. (b) The cascade window as a function of the ratio of senior loans to junior loans, $\langle l_\sen \rangle / \langle l_\jun \rangle$, achieves a minimum size at $\approx 1.79$.} 
\label{fig:cascaderegion}
\end{center}
\end{figure}

The blue region in Fig.~\ref{fig:cascaderegion}(a) is the multiplex cascade region, the set of average out-degrees $(\langle l_\jun \rangle, \langle l_\sen \rangle)$ such that the first-order cascade condition [inequality~\eqref{eq:FOcascadecondition}] is satisfied. Note that, as pointed out in~\cite[page 3]{Brummitt2012_PRER}, we have $\lambda_{\max}(\jacobian) \geq \max \{\jacobian_{\jun\jun}, \jacobian_{\sen\sen}\}$. Thus, a sufficient condition for the multiplex cascade condition to hold is that at least one kind of default $\alpha \in \{\jun, \sen\}$ is supercritical (meaning that $\jacobian_{\alpha \alpha} > 1$) even in the absence of the other kind of default. Thus, the blue region in Fig.~\ref{fig:cascaderegion}(a) contains both the orange and green regions.

Importantly, however, the blue region contains pairs of intermediate edge densities $(\langle l_\jun \rangle, \langle l_\sen \rangle)$ outside the junior-only and senior-only cascade regions. In this part of parameter space, neither junior-defaults nor senior-defaults alone suffice to cause global cascades (i.e., $\jacobian_{\jun \jun} < 1$ and $\jacobian_{\sen \sen} < 1$), but a large cascade of defaults can nevertheless occur. These cascades, akin to the ``multiplexity-facilitated cascades'' in~\cite{Brummitt2012_PRER,Lee2014}, result from the interaction of junior- and senior-defaults. 

\subsection{Optimal seniority ratio}
\label{sec:optimal_seniority_ratio}

Note that the multiplex cascade region, show in blue in Fig.~\ref{fig:cascaderegion}(a), is not symmetric with respect to the average out-degrees in the two layers, $\langle l_\jun \rangle$ and $\langle l_\sen \rangle$. This asymmetry of the cascade region results from the asymmetric role that junior and senior loans play in the response function [Eq.~\eqref{eq:define_thresholds}]. Given this asymmetry, what ratio of junior and senior loans is optimal in some sense?

Suppose that the average total degree $\langle l_\jun \rangle + \langle l_\sen \rangle$ is kept constant and that we tune the fraction of loans that are \new{junior}. For some loan densities $\langle l_\jun \rangle + \langle l_\sen \rangle$, the multiplex cascade condition~\eqref{eq:FOcascadecondition} holds no matter the fraction of \new{junior} loans. For example, the line segment $\{(\langle l_\jun \rangle, \langle l_\sen \rangle) : 4 = \langle l_\jun \rangle + \langle l_\sen \rangle\}$ is contained within the multiplex cascade region in Fig.~\ref{fig:cascaderegion}(a), so defaults spread widely no matter the composition of seniority levels. 
In other cases, such as the line segment of mean degrees satisfying $7 = \langle l_\jun \rangle + \langle l_\sen \rangle$ in Fig.~\ref{fig:cascaderegion}(a), the cascade condition~\eqref{eq:FOcascadecondition} holds if and only if the fraction of \new{junior} loans is \new{very small or somewhat large}. In cases like this one, having an intermediate fraction of \new{junior} debts prevents defaults from spreading widely.

Now suppose that we can set the ratio $\langle l_\sen \rangle / \langle l_\jun \rangle$ of the average numbers of senior and junior loans per bank, but the total density of loans $\langle l_\sen \rangle + \langle l_\sen \rangle$ can vary. For instance, the number of loans may increase during an economic boom and decrease during a recession, yet the fraction of loans that are senior does not change. Thus, we are considering rays through the origin in Fig.~\ref{fig:cascaderegion}(a). 

For ``seniority ratios'' $\sigma \geq 0$, define the \emph{cascade window} to be the set of average degrees $(\langle l_\sen \rangle, \langle l_\jun \rangle)$ such that $\langle l_\sen \rangle / \langle l_\jun \rangle = \sigma$ and such that the first-order cascade condition~\eqref{eq:FOcascadecondition} holds. Figure~\ref{fig:cascaderegion}(b) shows this cascade window as a function of the seniority ratio $\langle l_\sen \rangle / \langle l_\jun \rangle$. Note that the size of the cascade window is minimized at a unique value of $\langle l_\sen \rangle / \langle l_\jun \rangle$, which is approximately $1.79$ in Fig.~\ref{fig:cascaderegion}. 
We define the \emph{optimal seniority ratio} to be this minimizer, or more precisely,
\begin{subequations}
\begin{align}
&\frac{\langle l_\sen \rangle}{\langle l_\jun \rangle}\phantom{}_\text{optimal} := \argmin_{\sigma \geq 0} \vert \{ r \geq 0 : \lambda_{\max} (\jacobian) 
\geq 1 \} \vert \label{eq:define_optimal_ratio_top}
\end{align}
\new{where $r$ is the distance from the origin, $\vert \cdot \vert$ denotes Lebesgue measure on $\mathbb{R}$, and  $\lambda_{\max} (\jacobian)$ is evaluated at}
\begin{align}
\langle l_\jun \rangle = \frac{r}{\sqrt{1+\sigma^2}}, \quad \langle l_\sen \rangle = \frac{r \sigma }{ \sqrt{1+\sigma^2}},\label{eq:define_optimal_ratio_evaluation}
\end{align}
\label{eq:define_optimal_ratio}
\end{subequations}
\new{where we have changed from polar to Cartesian coordinates via $(r, \arctan \sigma) \mapsto (\langle l_\jun \rangle, \langle l_\sen \rangle)$.} 
%
%
%
If this optimal seniority ratio \new{$({\langle l_\sen \rangle}/{\langle l_\jun \rangle}\phantom{})_\text{optimal}$} is greater than one (as in Fig.~\ref{fig:cascaderegion}), then senior debts should be issued more than junior debts (on average) for the sake of minimizing the size of the set of network densities such that bankruptcies spread widely. That is, if a financial network keeps approximately the same ratio of senior to junior loans \new{($\langle l_\sen \rangle / \langle l_\jun \rangle$)} but changes the total number of loans (say, over the course of a business cycle), then there should be more senior loans than junior loans in order to minimize the risk of a large cascade of defaults. 

Intuitively, senior loans are stabilizing because they are less significant early in a cascade: a node in junior-default causes losses (of assets) for its lenders (in-neighbors) in the junior layer only. Thus, changing some loans from junior to senior makes large cascades less likely, and hence it shrinks the cascade window. However, not all loans should be senior, because that case is identical to all loans being junior, which has a dangerously large cascade window [look at the horizontal and vertical axes of Fig.~\ref{fig:cascaderegion}(a)]. In between those extremes lies an optimal ratio of senior to junior loans. That optimal ratio exceeds one [the dashed black $45$-degree line in Fig.~\ref{fig:cascaderegion}(a)] because senior loans are less contagious, in the sense that nodes in junior-default can still repay their senior debts. 

In Fig.~\ref{fig:compare_theory_simulation} of Appendix~\ref{sec:cascade_region_fixed_point}, we show that the three types of cascade regions shown in Fig.~\ref{fig:cascaderegion}(a) (junior-only, senior-only, and multiplex) closely agree with the fixed point of the recursion equations~\eqref{eq:recursion_vector} and with numerical simulations of the model \new{for uncorrelated \ER multiplex networks}. These comparisons also show that within the multiplex cascade region many nodes end up in junior-default, but many nodes are in senior-default only when senior loans greatly outnumber junior loans. In practice, the number of defaults of any type is of primary concern, so our definition of the optimal seniority ratio in Eq.~\eqref{eq:define_optimal_ratio} ignores types of defaults.

\subsection{Effect of varying the thresholds (ratio of equity to interbank assets)}

\begin{figure}
 \begin{center}
    \includegraphics{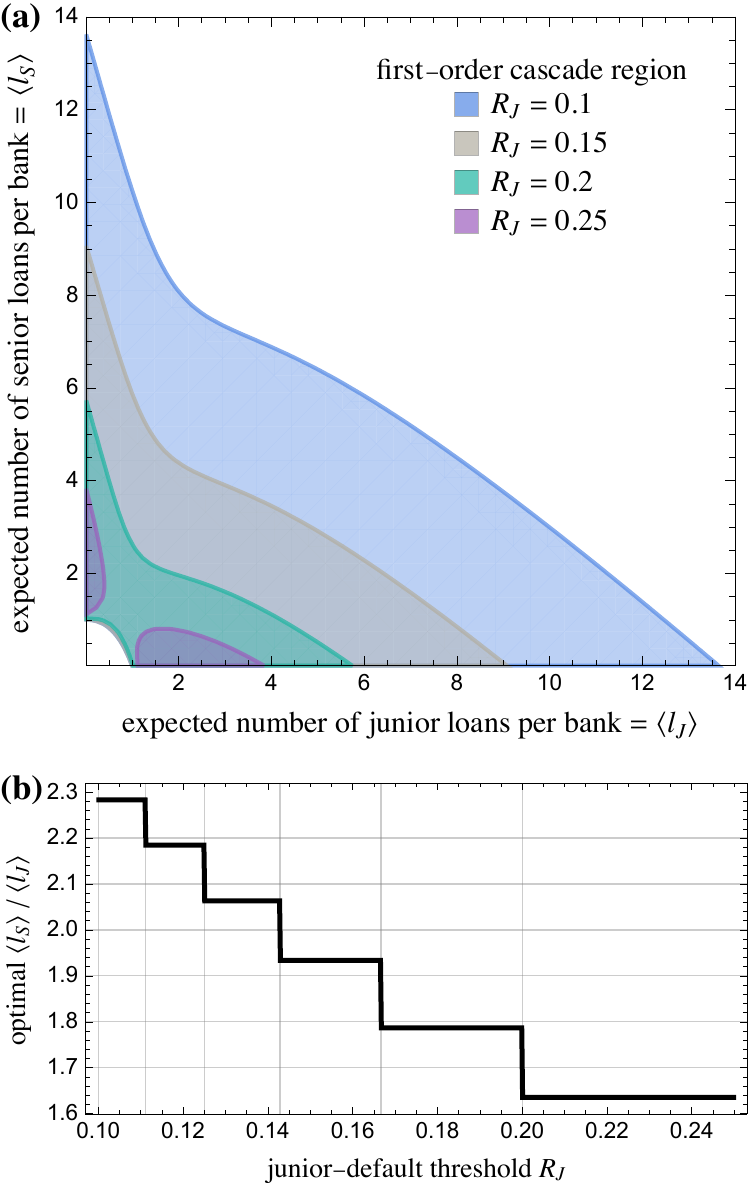}
  \caption{(Color online) 
  If banks hold less capital (equity) $w$ relative to their interbank assets $l_\jun + l_\sen$ (i.e., if the threshold for junior-default, $R_\jun$, decreases), then (a) the cascade region (the set of parameters satisfying the first-order cascade condition~\eqref{eq:FOcascadecondition}] enlarges, and (b) the optimal ratio of senior to junior loans increases. The first-order cascade condition~\eqref{eq:FOcascadecondition} considers defaults that occur due to one defaulted neighbor (for details, see Appendix~\ref{sec:derive_FOCC}). Thus, when $R_J$ falls below the inverse of an integer, $1/k$, nodes with total degree $k$ and with one defaulted neighbor will junior-default, so the cascade region changes at $R_J = 1/k$, and hence the optimal ratio jumps up at $R_J = 1/k$.
} 
  \label{fig:different_RJ}
 \end{center}
\end{figure}

How does the cascade region (and its optimal seniority ratio) change as the banks' threshold $R_\jun$ changes? Here we keep the assumption that banks all have the same junior-default threshold $R_\jun$, but we vary $R_\jun$. 
\new{The multiplex network is still a pair of uncorrelated \ER random graphs with mean degrees $\langle l_\jun \rangle$ and $\langle l_\sen \rangle$.} 
If banks have a smaller capital ratio $R_\jun$, then default becomes more likely, so the multiplex cascade region grows, as shown in Fig.~\ref{fig:different_RJ}(a). Conversely, as $R_\jun$ increases, the cascade region shrinks, and it can even splinter into two regions [see the purple regions for $R_\jun = 0.25$ in Fig.~\ref{fig:different_RJ}(a)]. In cases like this one with two disjoint cascade regions, the optimal ratio defined in Eq.~\eqref{eq:define_optimal_ratio} does not exist because the set 
\new{$\{ r \geq 0 : \lambda_{\max} (\jacobian) \geq 1 \}$}
is empty for some $\sigma$. [A sensible definition for the optimal seniority ratio 
could be the slope that maximizes the distance from the ray through the origin (with that slope) to the two cascade regions.]

The asymmetry between junior and senior loans also changes with the junior-default threshold $R_\jun$. In particular, the optimal seniority ratio increases as $R_\jun$ decreases, as shown in Fig.~\ref{fig:different_RJ}(b). 


In some financial crises, many banks find themselves in financial distress, because of losses in external assets, for instance. 
The model studied here  approximates this scenario as a small junior-default threshold $R_\jun$ for every bank (and a small fraction of banks initially in default). 
Decreasing $R_\jun$ enlarges the cascade region, as illustrated in Fig.~\ref{fig:different_RJ}(a). However, 
the banks can reduce systemic risk if they increase, on average, the fraction of debts that are senior. Figure~\ref{fig:different_RJ}(b) shows the optimal amount of that increase.


\section{More than two seniority levels}
\label{sec:M_layers}

Now we show that mixing among several seniority levels can further reduce systemic risk (in the sense of shrinking the cascade region).

We generalize the model to $M \geq 1$ seniority levels. Let $(1,2,\ldots ,M)$ denote the indices of the seniority levels in increasing order of seniority, and let $R_1$ denote the threshold for default at the most junior layer. The response function for default at level $i$, derived in Appendix~\ref{appendix:thresholds_for_default_many_seniority_levels}, is 
\begin{align} 
   F_{i}(\vec{l},\vec{b},\vec{m}) = \begin{cases} 1 & \text{if }
   \sum_{s=1}^{M}m_{s}-\sum_{k=1}^{i-1}b_{k} > R_{1} \sum_{s=1}^M l_{s}
   \\ 0 & \text{otherwise} \end{cases}.
   \label{eq:response_function_M_layers}
 \end{align}
By convention, sums of the form $\sum_{k=1}^0$ are zero.

\subsection{First-order cascade condition}
The cascade condition, a straightforward generalization of the one for $M=2$ layers in Sec.~\ref{sec:multiplex_cascade_condition}, is that the Jacobian matrix $\jacobian^M \in \mathbb{R}^{M \times M}$ has largest eigenvalue exceeding one. The $(i,j)$ entry of $\jacobian^M$ is 
\begin{subequations}
\begin{align}
 \jacobian_{ij}^M 
 &= \E{\!}_{M} \left [l_j \indicator{1 - \sum_{k = 1}^{i-1} b_{k} > R_1 \left (\sum_{\alpha = 1}^M l_\alpha \right )} \right ] \label{eq:Jij_first} \\
 &= \prod_{k=1}^{i-1}p_{0}^{k,\text{borrow}} \E{\!}_{M} \left [l_j \indicator{1 > R_1 \left (\sum_{\alpha = 1}^M l_\alpha \right )} \right ]  \label{eq:Jij_second} \\
 &= \prod_{k=1}^{i-1}p_{0}^{k,\text{borrow}} \sum_{\vec l : \sum_{\alpha = 1}^M l_\alpha < 1/R_1}\prod_{s=1}^{M}p_{l_s}^{s,\text{loan}} l_{j}.\label{eq:Jij_third}
\end{align}
\label{eq:Jij_written_a_few_ways}
\end{subequations}
Here, products of the form $\prod_{k=1}^{0}$ are defined to be $1$, and the expectations $\E_M$ are  over $p_{l_1}^{1, \text{loan}}, \ldots, p_{l_M}^{M,\text{loan}}$ and over $p_{b_1}^{1, \text{borrow}}, \ldots ,p_{b_{M-1}}^{M-1, \text{borrow}}$. The expression in~\eqref{eq:Jij_second} equals that in~\eqref{eq:Jij_first} because the threshold $R_1 > 0$, so the indicator in~\eqref{eq:Jij_first} is $1$ only if the bank has no liabilities of seniority $k$ (i.e., $b_k = 0$) for all $k=1, 2, \dots, i-1$; furthermore, because a bank's in- and out-degrees are independent, the indicator in Eq.~\eqref{eq:Jij_first} factors to $\indicator{b_k = 0 \,  \forall \, k \in \{1, 2, \dots, i-1\}} \indicator{1 > R_1 (\sum_{\alpha = 1}^M l_\alpha)}$. Equation~\eqref{eq:Jij_third} writes out the expectation in Eq.~\eqref{eq:Jij_second} by summing over $\vec l \in \{0, 1, 2, \dots\}^M$ such that $\sum_{\alpha = 1}^M l_\alpha < 1/R_1$. 

Observe from Eq.~\eqref{eq:Jij_second} that $\jacobian^M$ consists of rows that are linearly dependent. Thus, $\jacobian^M$ has rank one and only one non-zero eigenvalue. Because the trace of a matrix equals the sum of its eigenvalues, the largest eigenvalue of $\jacobian^M$ is its trace, which, from Eq.~\eqref{eq:Jij_third}, is given by
 \begin{eqnarray}
 \!\!\!\!\!\!\!\! \mathrm{tr}\jacobian^{M} =  \sum_{i=1}^M \prod_{k=1}^{i-1}p_{0}^{k,\text{borrow}} \!\! \sum_{{\vec l : \sum_{\alpha = 1}^M l_\alpha < 1/R_1}} \prod_{s=1}^{M}p_{l_s}^{s,\text{loan}}
  l_{i}.
  \label{eq:trJ}
 \end{eqnarray}
 Therefore, we can analytically express $\lambda_{\max}(\jacobian^{M})$ by using the degree distributions and the threshold value $R_1$ on the most junior layer. This expression immediately gives the first-order cascade condition, $\mathrm{tr}\jacobian^{M}>1$.

\subsection{Distinguishing more seniority levels reduces risk of large cascades of default}
\label{sec:M_ER_layers}

If we realize that not all loans have the same seniority level, how does vulnerability to large cascades change? For simplicity, we consider ``splitting'' an \ER random graph into $M$ independent, identically distributed layers (as in~\cite{Brummitt2012_PRER}). That is, we compare  a single-layer \ER random graph with mean out-degree $\meanDegreeInEachLayer$ to a multiplex \ER random graph with mean out-degree $\meanDegreeInEachLayer/M$ in each of the $M$ layers.

The Jacobian matrix $\jacobian^\text{$M$-ER}$ for an $M$-layer directed \ER random graph with edge density $\langle l_i \rangle$ in each layer $i \in \{1, 2, \dots, M\}$ is, from Eq.~\eqref{eq:Jij_written_a_few_ways}, 
\begin{align}
\jacobian_{ij}^\text{$M$-ER} &= \frac{\langle l_i \rangle \Gamma(\lceil 1/R_1 \rceil -1, \sum_{i=1}^M \langle l_i \rangle)}{\Gamma(\lceil 1/R_1 \rceil - 1)} \exp \left [ - \sum_{k=1}^{j-1} \langle l_k \rangle \right ],
\label{eq:Jacobian_M_ER}
\end{align}
where the incomplete gamma function $\Gamma \left (x, y \right ) \equiv \int_y^\infty u^{x-1} e^{-u} du$ and the gamma function $\Gamma (x) \equiv \Gamma (x, 0)$.

\begin{figure}[htb]
\includegraphics{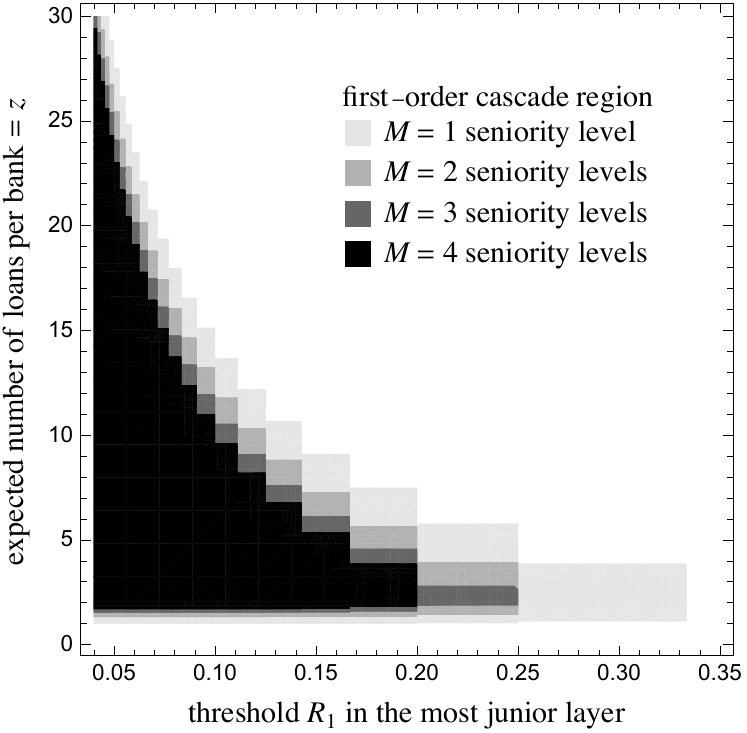}
\caption{Distinguishing more types of seniority reduces the size of the cascade region. In other words, with more types of seniority, the expected size of a default cascade is negligibly small for more pairs of thresholds $R_1$ and expected number of loans per bank $z = \sum_{i=1}^M \langle l_i \rangle$. Here we consider splitting a directed \ER network into $M \in \{1,2,3,4\}$ many independent, identically distributed layers, each with the same mean degree $\langle l_i \rangle = z / M$.}
\label{fig:split_into_four_layers}
\end{figure}

Figure~\ref{fig:split_into_four_layers} shows the resulting cascade region (i.e., the parameters satisfying $\mathrm{tr} \jacobian^\text{$M$-ER} > 1$) for $M \in \{1,2,3,4\}$ many \ER layers that each have mean out-degree $\langle l_i \rangle = \meanDegreeInEachLayer / M$. Here, the parameters are the threshold $R_1$ for default at the most junior level (horizontal axis) and the mean total out-degree $\meanDegreeInEachLayer$ (vertical axis). 

Notice in Fig.~\ref{fig:split_into_four_layers} that the cascade region shrinks with the number $M$ of seniority levels, meaning that the system is less vulnerable to large cascades of default. (By contrast, in the model in~\cite{Brummitt2012_PRER}, the cascade region \emph{grows} with $M$.) For some capital-to-interbank-asset ratios $R_1$, the cascade region is even eliminated as $M$ grows (e.g., $R_1=0.3$ for $M=1$).

The reason the cascade region shrinks with $M$ is that making a loan more senior makes it less significant early in a cascade. For instance, if $M=1$, then a defaulted node reduces the assets of all of its lenders (in-neighbors), whereas if $M > 1$, then only the in-neighbors in layer $1$ lose an asset, so less contagion occurs. The decline in importance of more senior layers is captured by the multiplicative factor $\prod_{k=1}^{i-1}p_{0}^{k,\text{borrow}}$ in Eq.~\eqref{eq:Jij_third} and by the exponential factor in Eq.~\eqref{eq:Jacobian_M_ER}. In short, distinguishing more types of seniority makes more interbank debts insignificant (early in a crisis); hence a systemic crisis is less likely.

This result hinges on the assumption that banks' buffers against insolvency $R_1$ remain constant as the number of seniority levels ($M$) grows. It is possible that if there were more seniority levels, then banks would erode their buffers against insolvency, $R_1$, thereby pushing the system back toward the cascade region, where bankruptcies likely spread widely.

\section{Heavy-tailed degree distributions\label{sec:heavytailed}}

\begin{figure}
 \begin{center}
 \includegraphics{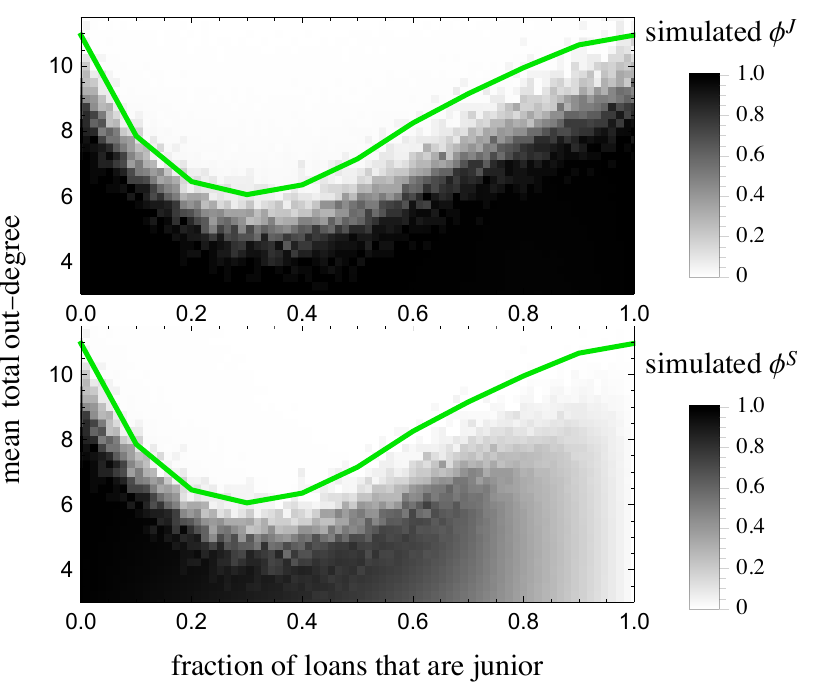}
 \caption{
\new{
(Color online) 
For heavy-tailed degree distributions like those in the Brazilian interbank lending network~\cite{Cont2013}, having $\approx 30\%$ of loans be junior (and the rest senior) reduces systemic risk, in that it minimizes the size of the set of mean total out-degrees such that bankruptcy spreads widely. 
The gray-scale background shows the average fraction of banks in junior-default (top panel) and in senior-default (bottom panel) in simulations on $2400$ banks, begun from $10$ banks initially in senior-default, averaged over $30$ simulations. The green lines mark the boundary of the cascade region [$\lambda_{\max}(\jacobian) = 1$ with $\jacobian$ computed from Eq.~\eqref{eq:J_factors}]. 
The junior-default threshold of every bank is $R_\jun = 0.18$, as in Fig.~\ref{fig:cascaderegion}. Directed graphs are generated with the static model~\cite{Goh2001,Catanzaro2005_EPJB} with out-degree distributions having tail behavior $\propto k^{-\gamma}$ with $\gamma = 2.83$, which approximates the value found in the Brazilian interbank lending network~\cite{Cont2013}. A fraction of loans (horizontal axis) chosen uniformly at random are junior, and the rest are senior. The ``optimal'' fraction of junior loans (that minimizes the height of the green line) is $\approx 0.30$. For the mean out-degree (vertical axis) of $\approx 8.5$ observed for Brazilian banks~\cite{Cont2013}, the expected cascade size is small if and only if an {intermediate} fraction of loans are junior.\\
}
}
\label{fig:staticmodel}
\end{center}
\end{figure}

\new{
Real networks of loans among banks show evidence of somewhat heavy-tailed degree distributions~\cite{Boss2004,DeMasi2006,Cont2013}. 
For instance, the network of Brazilian banks in 2007--2008~\cite{Cont2013} has in- and out-degree distributions that have tails approximated by a power law $\propto k^{-\gamma}$ with exponent $\gamma \approx 2.8$, and the mean in- and out-degree are both approximately $8.5$. 

To capture such heavy-tailed degree distributions, we use the static model of scale-free networks~\cite{Goh2001,Catanzaro2005_EPJB}, which allows tuning the mean degree and the exponent $\gamma$ of the degree distribution~\cite{footnoteOnStaticModel}. 
Because we lack access to data on seniority levels of loans, we first create a random graph with a certain mean out-degree, and then we assign a fraction of directed edges (chosen uniformly at random) to be junior, and the remaining are senior.

Figure~\ref{fig:staticmodel} shows that, in numerical simulations and in the analytical approximation~\eqref{eq:FOcascadecondition}, the fraction of loans that are junior can determine whether large cascades of defaults tend to occur. Notice the ``valley'' in Fig.~\ref{fig:staticmodel}: there is a narrow range of ``optimal'' values of the fraction of junior loans (horizontal axis of Fig.~\ref{fig:staticmodel}) that shrinks the set of mean-degrees such that cascade size is large. This optimal fraction of junior loans is approximately $0.30$, which translates to an optimal seniority ratio of $\langle l_\sen \rangle / \langle l_\jun \rangle \approx (1-0.30)/0.30 \approx 2.3$, which is qualitatively similar to the optimal seniority ratio $1.79$ found for multiplex \ER networks in Fig.~\ref{fig:cascaderegion}. 

Notice in Fig.~\ref{fig:staticmodel} that the empirical mean out-degree in the Brazilian network (approximately $8.5$)~\cite{Cont2013} intersects the black regions (where the mean cascade size is large) whenever the fraction of junior loans is large or small. We emphasize that we do not know the capital levels of these Brazilian banks, which affects the threshold for junior-default, $R_\jun$. However, Fig.~\ref{fig:staticmodel} suggests that seniority levels of loans in real-world networks could potentially affect whether default might spread widely or not.
}

\section{Conclusion and discussion}
\label{sec:conclusion}

We model seniority of debt using a multiplex network, and we study the likelihood of bankruptcy spreading widely using a generalized cascade condition. The model has interesting tradeoffs. Whereas self-interested banks would prefer to make their loans at the most senior level, a mixture of junior and senior debts optimally reduces systemic risk, in that it minimizes the \new{size of the} set of network densities such that cascades are likely large.  Such disparity between individual interests and system-wide stability is the ``regulator's dilemma"~\cite{Beale2011}. For two types of seniority, \new{and for edges placed uniformly at random (i.e., uncorrelated \ER networks) or according to a heavy-tailed degree distribution}, we show that the optimal multiplex networks have on the order of $50\%$ to $100\%$ more senior debts than junior debts. 
\new{To give theoretical support for policy suggestions, however, will require considering other factors that we have abstracted away, such as interest rates for different seniority levels, which affect the incentives of banks.}

\new{Traditionally, cascades of losses in financial networks have been studied using the Eisenberg-Noe algorithm~\cite{Eisenberg2001} for computing a market-clearing payment vector (with losses shared proportionally),
which was recently extended to multiple seniority levels~\cite{Gourieroux2013}. 
The Eisenberg-Noe algorithm is economically meaningful in normal times, but as Cont et al.~\cite{Cont2013} point out, the algorithm is not so reasonable for modeling an extreme event such as the default of a large bank. We instead take an approach typical in physics of studying ensembles of financial networks~\cite{Watts2002,GaiKapadia2010}. We show that cascades on networks with different seniority may be analyzed with a variant of multiplex threshold cascade models 
studied in physics~\cite{Brummitt2012_PRER,Yagan2012,Lee2014}.}

The seniority of debt is just one dimension of multiplexity in financial networks. Other dimensions of debt include the maturity of loans and whether loans are collateralized~\cite{Bargigli2015,Montagna2013}. 
Furthermore, other important interactions include the trade of financial assets~\cite{Brummitt2014} and common exposures (i.e., multiple institutions holding the same financial assets)~\cite{Montagna2013,Caccioli2015}.

\new{We call our model ``multilevel contagion'' because a node can become ``contagious'' in one layer without necessarily becoming contagious on another layer (e.g., a bank can be in junior-default but not in senior-default). By contrast, in other models of threshold contagion in multiplex networks~\cite{Brummitt2012_PRER,Yagan2012,Lee2014} and in variants of percolation on multiplex networks~\cite{Buldyrev2010,Son2012,Baxter2012,Cellai2013,Min2014_robustness_multiplex_interlayer_degree_correlations}, when a node changes state in one layer, it necessarily changes state in other layers. Multilevel contagion might give insight into other kinds of social contagion. For example, 
one might use it to model the adoption of a new product if ``fans'' of the product spread influence to neighbors of one type but ``super-fans'' spread influence to neighbors of two types (a multiplex version of the multistage complex contagion model~\cite{Melnik2013}). 
We hope our model will stimulate further research on these many types of cascades in social, financial, and other multiplex networks.
}

\section*{Acknowledgments}
C.D.B. is supported by the James S. McDonnell Postdoctoral Fellowship in Studying Complex Systems. 
T.K. is supported by KAKENHI 25780203 and 24243044.
The authors thank the Isaac Newton Institute for Mathematical Sciences (Cambridge, UK) for support and hospitality during the programme Systemic Risk: Mathematical Modelling and Interdisciplinary Approaches, where work on this paper was undertaken.

\onecolumngrid   
\appendix 




\section{Thresholds of default for $M$ levels of seniority}
\label{appendix:thresholds_for_default_many_seniority_levels}
Here we consider the general case of $M \geq 1$ levels of seniority of interbank debt (rather than just two, junior and senior), and we label the levels of seniority by $\alpha \in \{1, 2, \dots, M\}$. We choose the convention that debts of seniority $\alpha$ are more senior than debts of seniority $\alpha'$ whenever $\alpha > \alpha'$.

Recall from Fig.~\ref{fig:balance_sheet} the stylized balance sheet of a bank: a bank's assets consist of external assets $e$ and $l_\alpha$ many interbank loans of seniority $\alpha \in \{1, 2, \dots, M\}$, while its liabilities consist of external liabilities $d$ and $b_\alpha$ many interbank liabilities of seniority $\alpha \in \{1, 2, \dots, M\}$. We assume that the external liabilities $d$ (which could include, for example, demand deposits) are more senior than debts to other financial institutions. The equity $w$ is the difference between assets and liabilities, $$w = e - d + \sum_{\alpha = 1}^M (l_\alpha - b_\alpha).$$

Now suppose that this bank has some interbank assets (i.e., $\sum_{\alpha = 1}^M l_\alpha > 0$) and that it suffers losses on $m_\alpha$ of its $l_\alpha$ many interbank assets (for $\alpha \in \{1, 2, \dots, M\}$), so that its interbank assets now have total value $\sum_{\alpha = 1}^M (l_\alpha - m_\alpha)$. 
Then exactly one of the following events occurs:
\begin{itemize}
\item The bank is  solvent if and only if the capital buffer $w$ can absorb the losses, i.e., $w \geq \sum_{\alpha=1}^M m_\alpha$.
\item The bank is in default at layer $\alpha' \in \{1, 2, \dots, M\}$ but not in any layer $\geq \alpha' + 1$ if and only if 
\begin{align}
w + \sum_{\alpha = 1}^{\alpha' - 1} b_{\alpha}
< 
\sum_{\alpha = 1}^M m_{\alpha}
\leq 
w + \sum_{\alpha = 1}^{\alpha'} b_{\alpha}.
\label{eq:default_layer_alpha}
\end{align}
\item The bank is in \emph{complete default}, meaning that it cannot pay off any of its interbank debts nor any of its liabilities to external creditors, if and only if 
$$\sum_{\alpha = 1}^M m_\alpha > w + \sum_{\alpha = 1}^M b_{\alpha}.$$
\end{itemize}
Here, we focus on cascades in the interbank network; for this purpose, we do not need to distinguish between default at the most senior level ($M$) from complete default because both have the same effect on the banks that had lent to the defaulted bank. 

Rearranging the strict inequality in Eq.~\eqref{eq:default_layer_alpha} and dividing by the total lending $\sum_{\alpha = 1}^M l_{\alpha} > 0$ gives the response function in Eq.~\eqref{eq:response_function_M_layers}.

\section{Relationship with ``multistage complex contagion''}
\label{sec:relationship_with_multistage}
Our model is related to the ``multistage complex contagion" model of Melnik et al.~\cite{Melnik2013}. In their model,  
nodes exist in one of three states: inactive, active, and hyper-active. Their hyper-active nodes are active nodes with greater influence on their neighbors. To see the commonality between the models, consider our model with every node having the same junior threshold $R_\jun$ and with the multiplex network consisting of two identical layers (i.e., all the multiplex edges are ``fully overlapped''~\cite{Bianconi2013,Cellai2013}). Then every neighbor in junior-default (respectively, in senior-default) contributes $m_\jun + m_\sen = 1$ (respectively, $m_\jun + m_\sen = 2$) in the numerator of the response function. Thus, this model is equivalent to a special case of the model in~\cite{Melnik2013} in which i) the ``bonus influence parameter'' $\beta$ (which captures the extra influence of hyper-active nodes) takes the value $1$; ii) the network is a \emph{directed} single-layer graph; and iii) the thresholds for activation and hyper-activation are $R_\jun$ and $R_\jun + b_\jun / (l_\jun + l_\sen) = R_\jun + b_\jun / (2 l_\jun)$, respectively. To model real financial systems, however, edges should be only partially overlapped, 
\new{a phenomenon studied in the contexts of statistical mechanics in~\cite{Bianconi2013} and of percolation in~\cite{Cellai2013}.}
\new{Our multilevel contagion model could be generalized to such a partially overlapped case by adapting techniques from the study of mutual percolation on multiplex networks with arbitrary amounts of overlap~\cite{Cellai2013}, a nontrivial task left for future work.}

\section{Derivation of the first-order cascade condition}
\label{sec:derive_FOCC}
Now we obtain a simple expression that approximately captures for which parameters a vanishingly small seed leads to a cascade that results in a finite fraction of banks in junior- or senior-default. Note that if the threshold $R_\jun > 0$, then $(0,0)$ is a fixed point of the recursion~\eqref{eq:recursion_vector} [because $B_m^k(0) \equiv \indicator{m=0}$ and $F_\alpha(\vec l, \vec b, \vec 0) \equiv 0$ provided that $R_\jun > 0$]. The \emph{first-order cascade condition} is the linear instability of Eq.~\eqref{eq:recursion_vector} at this fixed point at the origin $(\phi_t^\jun, \phi_t^\sen) = (0,0)$ in the double limit $\phi_0^\jun \to 0$ and $\phi_0^\sen \to 0$. This linear instability provides a sufficient condition for the numbers of junior- and senior-defaulted banks to grow (at least initially). The Jacobian matrix of the right-hand side of Eq.~\eqref{eq:recursion_vector} is
\begin{align}
\jacobian &= \begin{pmatrix} \jacobian_{\jun\jun} & \jacobian_{\jun\sen} \\ \jacobian_{\sen\jun} & \jacobian_{\sen\sen} \end{pmatrix} \\
&= \begin{pmatrix} \E [ l_\jun (\indicator{1 > R_\jun (l_\jun + l_\sen)}-\indicator{0 > R_\jun})] & \E [ l_\sen (\indicator{1 > R_\jun (l_\jun + l_\sen)} - \indicator{0 > R_\jun})] \nonumber\\ \E [ l_\jun (\indicator{1-b_\jun > R_\jun (l_\jun + l_\sen)} - \indicator{-b_\jun > R_\jun (l_\jun + l_\sen)})] & \E[l_\sen (\indicator{1-b_\jun > R_\jun (l_\jun + l_\sen)} - \indicator{-b_\jun > R_\jun (l_\jun + l_\sen)})]  \end{pmatrix} \label{eq:J_complicated} \\
\end{align}
(The expectations $\E$ are over the degree distributions $p_{l_\jun}^{\jun,\text{loan}}, p_{l_\sen}^{\sen,\text{loan}}, p_{b_\jun}^{\jun,\text{borrow}}$.) Assume that $R_\jun > 0$, so that the factors $\indicator{0>R_\jun}$ in the top row of~\eqref{eq:J_complicated} vanish. Also, note that $b_\jun \geq 0$ and that $R_\jun > 0$ and $l_\jun+l_\sen\geq 0$, so the factors $\indicator{-b_\jun > R_\jun (l_\jun + l_\sen)}$ in the bottom row of Eq.~\eqref{eq:J_complicated} vanish. Finally, the other indicator in the bottom row of Eq.~\eqref{eq:J_complicated}, namely $\indicator{1-b_\jun > R_\jun (l_\jun + l_\sen)}$, simplifies to $\indicator{b_\jun = 0} \indicator{1 > R_\jun (l_\jun + l_\sen)}$ because $b_\jun \in \{0, 1, 2, \dots\}$. Because each node's in- and out-degrees $(l_\jun, b_\jun, l_\sen, b_\sen)$ are all independent, we can factor the expectation $\E \left [ l_\alpha \indicator{b_\jun = 0} \indicator{1>R_\jun(l_\jun + l_\sen)} \right ] = p_{0}^{\jun, \text{borrow}} \E \left [ l_\alpha \indicator{1>R_\jun(l_\jun + l_\sen)} \right ]$ for both $\alpha \in \{\jun, \sen\}$ (i.e., for the bottom-left and bottom-right entries of Eq.~\eqref{eq:J_complicated}). It follows that 
\begin{align}
\jacobian &= \begin{pmatrix} \E [ l_\jun \indicator{1 > R_\jun (l_\jun + l_\sen)}] & \E [ l_\sen \indicator{1 > R_\jun (l_\jun + l_\sen)}] \\ p_{0}^{\jun, \text{borrow}} \E [ l_\jun \indicator{1 > R_\jun (l_\jun + l_\sen)}] & p_{0}^{\jun, \text{borrow}} \E[l_\sen \indicator{1 > R_\jun (l_\jun + l_\sen)}]  \end{pmatrix} \label{eq:J_simplified}\\
&= \E [ l_\jun \indicator{1 > R_\jun (l_\jun + l_\sen)}] \begin{pmatrix} 1 & 0 \\ p_{0}^{\jun, \text{borrow}} & 0 \end{pmatrix} + \E [ l_\sen \indicator{1 > R_\jun (l_\jun + l_\sen)}] \begin{pmatrix} 0 & 1 \\ 0 & p_{0}^{\jun, \text{borrow}} \end{pmatrix}. \label{eq:J_factors}
\end{align}
Note that $\det{\jacobian}=0$. Thus, from Eq.~\eqref{eq:FOcascadecondition}, the largest eigenvalue of the Jacobian is $\lambda_{\max}(\jacobian) = \mathrm{tr}\jacobian = \jacobian_{\jun\jun} +  \jacobian_{\sen\sen}$.
 
It should be pointed out that this equality, $\lambda_{\max}(\jacobian) = \mathrm{tr}\jacobian$, does not necessarily hold true in other variants of multiplex contagion models. For example, in multiplex models with (independent) undirected networks and/or an exogenous senior-default threshold $R_\sen$, one would generally have $\lambda_{\max}(\jacobian) \neq \mathrm{tr}\jacobian$. A more detailed explanation is given in Appendix~\ref{sec:alternative_multiplex_models}.  

\section{Comparison between the first-order cascade condition with the fixed point of the recursion equations and with numerical simulations}
\label{sec:cascade_region_fixed_point}

\begin{figure*}
 \begin{center}
 \includegraphics{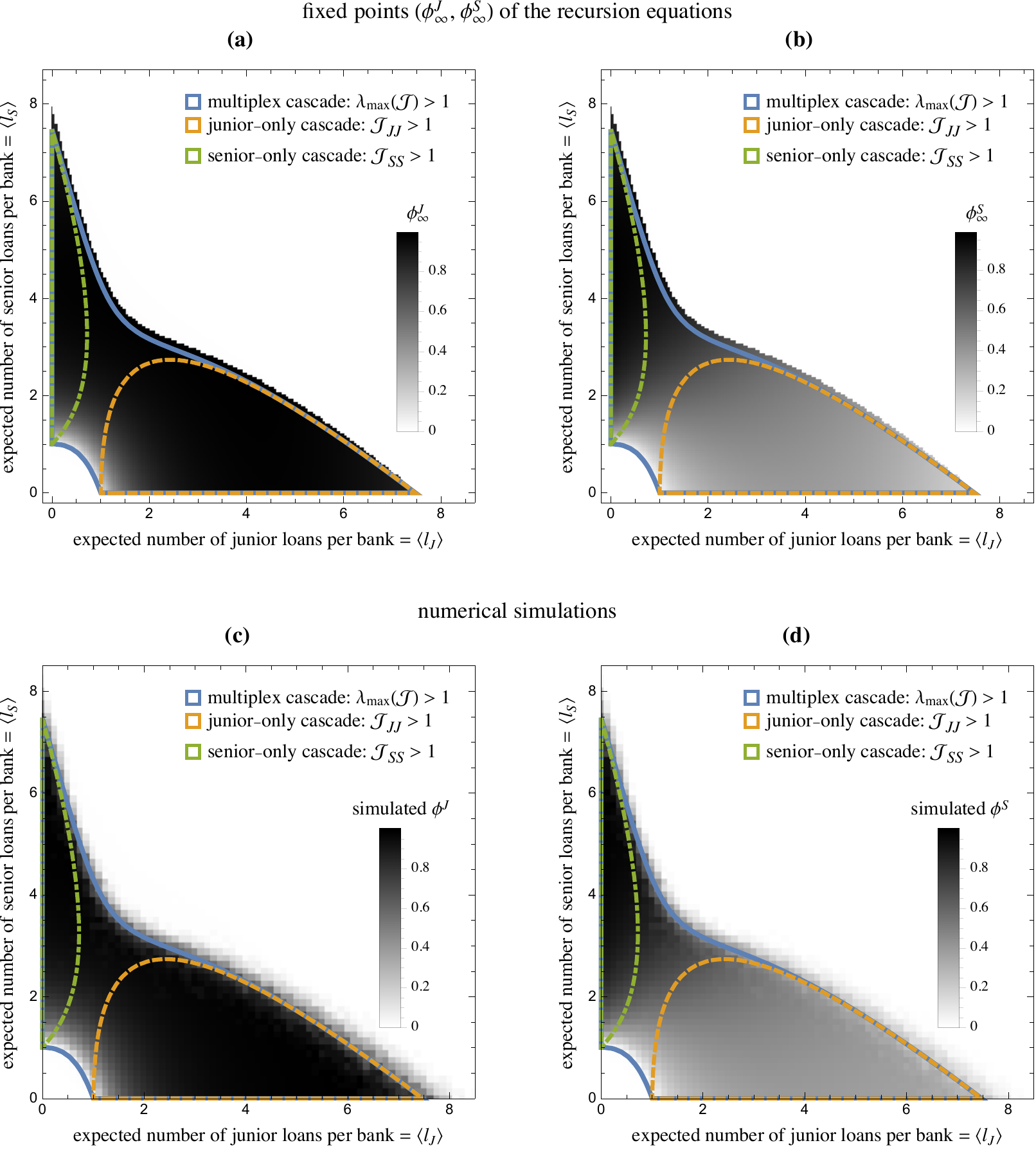}
 \caption{(Color online) 
Here we compare the junior-only, senior-only and multiplex cascade conditions [Eqs.~\eqref{eq:FOcascadecondition},\ \eqref{eq:junior_only_cascade_condition}, and\ \eqref{eq:senior_only_cascade_condition}, respectively] with the fixed points of the recursion equations~\eqref{eq:recursion_vector} [in panels (a) and (b)] and with numerical simulations on random graphs with $10^4$ nodes [in panels (c) and (d)]. 
The junior-default threshold of every node is $R_\jun = 0.18$ (as in Fig.~\ref{fig:cascaderegion}). 
The axes are the edge densities $\langle l_\jun \rangle$ and $\langle l_\sen \rangle$ (i.e, the average numbers of junior and senior loans per bank). The solid blue, dashed orange, and dot-dashed green lines 
surround the multiplex, junior-only, and senior-only cascade regions [i.e., the sets of values of $(\langle l_\jun \rangle, \langle l_\sen \rangle)$ that satisfy the inequalities\ \eqref{eq:FOcascadecondition},\ \eqref{eq:junior_only_cascade_condition}, and\ \eqref{eq:senior_only_cascade_condition}, respectively]. 
In panels (a) and (b), the grayscale in the background shows the 
fixed points $\phi_\infty^\jun$ and $\phi_\infty^\sen$ of the  recursion~\eqref{eq:recursion_vector}, begun from the initial condition $\phi_0^\jun = \phi_0^\sen = 5 \times 10^{-4}$. 
In panels (c) and (d), the grayscale in the background shows the fraction of nodes in junior- and senior-default, respectively, in numerical simulations with $10^4$ nodes and $5$ banks initially in senior-default, averaged over $75$ simulations.  
In the upper-left part of all four panels, where most loans are senior, the fractions  of nodes $(\phi_\infty^\jun,\phi_\infty^\sen)$ in junior- and senior-default at the end of the cascade are both $\approx 1$. By contrast, in the lower-right part, where most loans are junior, the fraction $\phi_\infty^\jun$ of nodes in junior-default is large [$\phi_\infty^\jun \approx 1$; panels (a) and (c)], whereas the fraction $\phi_\infty^\sen$ of nodes in senior-default is $\approx 0.4$ [see the light-gray region surrounded by the dashed orange line in panels (b) and (d)]. 
}
\label{fig:compare_theory_simulation}
\end{center}
\end{figure*}

Figure~\ref{fig:compare_theory_simulation} shows that the first-order cascade condition [the parameters satisfying Eq.~\eqref{eq:FOcascadecondition}, which are enclosed by the solid blue line in Fig.~\ref{fig:compare_theory_simulation}] agrees closely with the fixed point $(\phi_\infty^\jun, \phi_\infty^\sen)$ of the recursion equations~\eqref{eq:recursion_vector} [Fig.~\ref{fig:compare_theory_simulation}(a) and~\ref{fig:compare_theory_simulation}(b)] and agrees reasonably closely with the results of numerical simulations [Fig.~\ref{fig:compare_theory_simulation}(c) and~\ref{fig:compare_theory_simulation}(d)]. 
In the recursion equations and simulations, we use the same initial condition, with $\phi_0^\jun = \phi_0^\sen = 5 \times 10^{-4}$. The threshold for junior-default is $R_\jun = 0.18$, as in Fig.~\ref{fig:cascaderegion}.  

Notice in Fig.~\ref{fig:compare_theory_simulation} that when a large cascade occurs, many nodes are in junior-default, but many nodes are in senior-default if and only if, loosely speaking, there are many more senior loans than junior ones. [Look at the dark region in the left-hand sides of Figs.~\ref{fig:compare_theory_simulation}(b) and~\ref{fig:compare_theory_simulation}(d).] In practice, we are most concerned with whether the number of bankruptcies of any type is large. Thus, our definition of the optimal seniority ratio in Sec.~\ref{sec:optimal_seniority_ratio} considered only the multiplex cascade region and not the fractions $\phi_\infty^\jun$ and $\phi_\infty^\sen$. 

\section{Multilevel cascades under alternative settings}
\label{sec:alternative_multiplex_models}

In Appendix~\ref{sec:derive_FOCC}, we claimed that the equality $\lambda_{\max}(\jacobian) = \mathrm{tr}\jacobian$, which is proved in the main text for our model, does not necessarily hold true in other variants of the multilevel cascade model. In what follows, we substantiate this claim by examining two variants of the 2-layer multiplex model.
 
\subsection{Exogenous senior thresholds}

 In the model described in the main text, the threshold of senior-default, $R_\sen$, depends on the threshold of junior-default, $R_\jun$, and on the in-degree of the node on the junior layer, $b_{\jun}$. In this subsection, we consider an alternative model of multilevel cascades in which the threshold of senior-default is an exogenous parameter. Let the (locally treelike) graph hang down like a tree from a root node chosen uniformly at random. 
 The recursion equation for the probability of default of a node $t$ hops above the leaves of the tree is 
 \begin{align}
\phiex_{t+1}^\alpha &= \hat{g}^{(\alpha)}(\phiex_t^{\jun}, \phiex_t^{\sen}) 
\equiv \phiex_0^\alpha + (1-\phiex_0^\alpha) \sum_{l_\jun + l_\sen \geq 1}  p_{l_\jun}^{J,\text{out}} p_{l_\sen}^{\sen,\text{out}}   
  \sum_{m_\jun = 0}^{l_\jun} \sum_{m_\sen = 0}^{l_\sen} B_{m_\jun}^{l_\jun}(\phiex_t^\jun) B_{m_\sen}^{l_\sen}(\phiex_t^\sen) \hat F_\alpha(\vec l, \vec m)
\end{align}
for $\alpha \in \{\jun, \sen\}$, where $p_{l_\jun}^{\jun,\text{out}}$ and $p_{l_\sen}^{\sen, \text{out}}$ denote the out-degree distributions on the junior and senior layers, respectively. The response functions are now $\hat F_{\alpha}(\vec{l},\vec{m}) = \indicator{{m_{\jun}+m_{\sen}} > R_{\alpha} (l_{\jun}+l_{\sen})}$ for $\alpha  \in \{\jun, \sen\}$, where  $0 < R_{\jun} \leq R_{\sen} \leq 1$.
 
The first-order cascade condition is thus given by $\lambda_{\max}(\hat\jacobian)>1$, where the entries of $\hat \jacobian$ are $\E [ l_j \indicator{1 > R_i (l_\jun + l_\sen)}]$ for $(i,j) \in \{\jun, \sen\}^2$. Notice that the equality $\det\hat\jacobian = 0$ does not necessarily hold if $R_\jun < R_\sen$. It follows that $\lambda_{\max} (\hat\jacobian)$ is not in general equal to $\mathrm{tr}\hat\jacobian$. 

\subsection{Undirected edges}

Next, consider a multilevel cascade model in which edges are undirected. Again, let the graph hang down like a tree from a root node chosen uniformly at random. Consider a node located $t$ hops above the leaves of the tree. The probability that this node is in $\alpha$-default due to its children, conditioned on its parent in layer $\alpha \in \{\jun, \sen\}$ being solvent at layer $\alpha$, is given by the recursion equations 
\begin{subequations}
\begin{align}
\thetaun_{t+1}^\jun &= \thetaun_0^\jun + (1-\thetaun_0^\jun) \sum_{l_\jun + l_\sen \geq 1}  \frac{p_{l_\jun}^{\jun}l_\jun}{\langle l_\jun \rangle} p_{l_\sen}^{\sen}  
 \sum_{m_\jun = 0}^{l_\jun} \sum_{m_\sen = 0}^{l_\sen} B_{m_\jun}^{l_\jun}(\thetaun_t^\jun) B_{m_\sen}^{l_\sen}(\thetaun_t^\sen) \bar F_\jun(\vec l, \vec m), \label{eq:thetaJ}\\
 \thetaun_{t+1}^\sen &= \thetaun_0^\sen + (1-\thetaun_0^\sen) \sum_{l_\jun + l_\sen \geq 1} p_{l_\jun}^{\jun}   \frac{p_{l_\sen}^{\sen}l_\sen}{\langle l_\sen \rangle}  
 \sum_{m_\jun = 0}^{l_\jun} \sum_{m_\sen = 0}^{l_\sen} B_{m_\jun}^{l_\jun}(\thetaun_t^\jun) B_{m_\sen}^{l_\sen}(\thetaun_t^\sen) \bar F_\sen(\vec l, \vec m),\label{eq:thetaS}
\end{align}
\end{subequations}
where one uses Eq.~\eqref{eq:thetaJ} [respectively, Eq.~\eqref{eq:thetaS}] if the node's parent is a neighbor in layer $\jun$ (respectively, layer $\sen$), and $p_{l_\alpha}^{\alpha}$ is the degree distribution on layer $\alpha$. Because the graph is unweighted, and because this node lies at the end of an $\alpha$-type edge chosen uniformly at random, its degree in layer $\alpha$ is the excess degree distribution, $p_{l_\alpha}^\alpha l_\alpha / \langle l_\alpha \rangle$. The response functions are
\begin{subequations}
 \begin{align} 
   \bar F_{\jun}(\vec{l},\vec{m}) &= \begin{cases} 1 & \text{if } \frac{m_{\jun}+m_{\sen}}{l_{\jun}+l_{\sen}} > R_{\jun} \\ 0 & \text{otherwise} \end{cases}, \label{eq:FJ_undirected}\\
   \bar F_{\sen}(\vec{l},\vec{m}) &= \begin{cases} 1 & \text{if } \frac{m_{\jun}+m_{\sen}-l_{\jun} }{l_{\jun}+l_{\sen}} > R_{\jun} \\ 0 & \text{otherwise} \end{cases}.\label{eq:FS_undirected}
 \end{align}
 \label{eq:resonse_functions_unweighted}
 \end{subequations}
Because the graphs are undirected, the term $-b_\jun$ in $F_\sen$ in the original model [Eq.~\eqref{eq:define_FS}] is now $-l_\jun$ in Eq.~\eqref{eq:FS_undirected}. 

The corresponding Jacobian matrix is 
 \begin{align}
\bar\jacobian 
&= \begin{pmatrix} \E_{\jun} [ (l_\jun -1)\indicator{1 > R_\jun (l_\jun + l_\sen)}] & \E_{\jun} [ l_\sen \indicator{1 > R_\jun (l_\jun + l_\sen)}] \\ p_{0}^{\jun}\E_{\sen} [ l_\jun \indicator{1 > R_\jun (l_\jun + l_\sen)}] & p_{0}^{\jun}\E_{\sen}[(l_\sen-1) \indicator{1 > R_\jun (l_\jun + l_\sen)}]  \end{pmatrix}, \label{eq:J_simplifiedUn}
\end{align}
where the expectations $\E_\jun$ and $\E_\sen$ are over the joint probability distributions $\frac{p_{l_\jun}^{\jun}l_\jun}{\langle l_\jun\rangle} p_{l_\sen}^{\sen}$ and  $p_{l_\jun}^{\jun}   \frac{p_{l_\sen}^{\sen}l_\sen}{\langle l_\sen\rangle}$, respectively. It follows that $\det\bar\jacobian$ is generally nonzero, and therefore $\lambda_{\max}(\bar\jacobian) = \mathrm{tr}\bar\jacobian$ does not necessarily hold. 

\section*{References}


\begin{thebibliography}{10}

\bibitem{GaiKapadia2010}
P~Gai and S~Kapadia.
\newblock {Contagion in financial networks}.
\newblock {\em Proceedings of the Royal Society A: Mathematical, Physical and
  Engineering Sciences}, 466:2401--2423, June 2010.
\newblock \href {http://dx.doi.org/10.1098/rspa.2009.0410}
  {\path{doi:10.1098/rspa.2009.0410}}.

\bibitem{Nier2007}
Erlend Nier, Jing Yang, Tanju Yorulmazer, and Amadeo Alentorn.
\newblock Network models and financial stability.
\newblock {\em Journal of Economic Dynamics and Control}, 31(6):2033--2060,
  2007.
\newblock URL:
  \url{http://www.sciencedirect.com/science/article/pii/S0165188907000097},
  \href {http://dx.doi.org/10.1016/j.jedc.2007.01.014}
  {\path{doi:10.1016/j.jedc.2007.01.014}}.

\bibitem{Kobayashi2014}
T.~Kobayashi.
\newblock {A model of financial contagion with variable asset returns may be
  replaced with a simple threshold model of cascades}.
\newblock {\em Economics Letters}, 124:113--116, 2014.
\newblock \href {http://dx.doi.org/10.1016/j.econlet.2014.05.003}
  {\path{doi:10.1016/j.econlet.2014.05.003}}.

\bibitem{Gai2011}
P~Gai, A~Haldane, and S~Kapadia.
\newblock {Complexity, concentration and contagion}.
\newblock {\em Journal of Monetary Economics}, 58:453--470, July 2011.
\newblock \href {http://dx.doi.org/10.1016/j.jmoneco.2011.05.005}
  {\path{doi:10.1016/j.jmoneco.2011.05.005}}.

\bibitem{Upper2011}
C.~Upper.
\newblock {Simulation methods to assess the danger of contagion in interbank
  markets}.
\newblock {\em Journal of Financial Stability}, 7:111--125, August 2011.
\newblock \href {http://dx.doi.org/10.1016/j.jfs.2010.12.001}
  {\path{doi:10.1016/j.jfs.2010.12.001}}.

\bibitem{May2010}
R.~M. May and N.~Arinaminpathy.
\newblock {Systemic risk: the dynamics of model banking systems}.
\newblock {\em Journal of the Royal Society Interface}, 7:823--838, May 2010.
\newblock \href {http://dx.doi.org/10.1098/rsif.2009.0359}
  {\path{doi:10.1098/rsif.2009.0359}}.

\bibitem{Hurd2013}
T~R Hurd and J.P. Gleeson.
\newblock {On Watts' cascade model with random link weights}.
\newblock {\em Journal of Complex Networks}, 1(1):25--43, May 2013.
\newblock \href {http://dx.doi.org/10.1093/comnet/cnt003}
  {\path{doi:10.1093/comnet/cnt003}}.

\bibitem{Bargigli2015}
L.~Bargigli, G.~di~Iasio, L.~Infante, F.~Lillo, and F.~Pierobon.
\newblock The multiplex structure of interbank networks.
\newblock {\em Quantitative Finance}, 15(4):673--691, 2015.
\newblock \href {http://dx.doi.org/10.1080/14697688.2014.968356}
  {\path{doi:10.1080/14697688.2014.968356}}.

\bibitem{Montagna2013}
M.~Montagna and C.~Kok.
\newblock {Multi-layered Interbank Model for Assessing Systemic Risk}.
\newblock {\em Kiel Working Papers no.1873}, September 2013.

\bibitem{Elliott2014}
Matthew Elliott, Benjamin Golub, and Matthew~O. Jackson.
\newblock Financial networks and contagion.
\newblock {\em American Economic Review}, 104(10):3115--53, 2014.
\newblock URL:
  \url{http://www.aeaweb.org/articles.php?doi=10.1257/aer.104.10.3115}, \href
  {http://dx.doi.org/10.1257/aer.104.10.3115}
  {\path{doi:10.1257/aer.104.10.3115}}.

\bibitem{Kivela2014_multilayer_review}
Mikko Kivel\"{a}, Alex Arenas, Marc Barthelemy, James~P. Gleeson, Yamir Moreno,
  and Mason~A. Porter.
\newblock Multilayer networks.
\newblock {\em Journal of Complex Networks}, 2(3):203--271, 2014.
\newblock \href {http://dx.doi.org/10.1093/comnet/cnu016}
  {\path{doi:10.1093/comnet/cnu016}}.

\bibitem{Boccaletti2014}
S.~Boccaletti, G.~Bianconi, R.~Criado, C.I. del Genio,
  J.~G\'{o}mez-Garde$\tilde{\text{n}}$es, M.~Romance,
  I.~Sendi$\tilde{\text{n}}$a-Nadal, Z.~Wang, and M.~Zanin.
\newblock The structure and dynamics of multilayer networks.
\newblock {\em Physics Reports}, 544(1):1--122, 2014.
\newblock URL:
  \url{http://www.sciencedirect.com/science/article/pii/S0370157314002105},
  \href {http://dx.doi.org/10.1016/j.physrep.2014.07.001}
  {\path{doi:10.1016/j.physrep.2014.07.001}}.

\bibitem{Buldyrev2010}
Sergey~V Buldyrev, Roni Parshani, Gerald Paul, H~Eugene Stanley, and Shlomo
  Havlin.
\newblock {Catastrophic cascade of failures in interdependent networks}.
\newblock {\em Nature}, 464(7291):1025--1028, April 2010.
\newblock \href {http://dx.doi.org/10.1038/nature08932}
  {\path{doi:10.1038/nature08932}}.

\bibitem{Son2012}
Seung-Woo Son, Golnoosh Bizhani, Claire Christensen, Peter Grassberger, and
  Maya Paczuski.
\newblock {Percolation theory on interdependent networks based on epidemic
  spreading}.
\newblock {\em EPL (Europhysics Letters)}, 97(1):16006, January 2012.
\newblock \href {http://dx.doi.org/10.1209/0295-5075/97/16006}
  {\path{doi:10.1209/0295-5075/97/16006}}.

\bibitem{Baxter2012}
G~J Baxter, S~N Dorogovtsev, A~V Goltsev, and J~F~F Mendes.
\newblock {Avalanche Collapse of Interdependent Networks}.
\newblock {\em Physical Review Letters}, 109(24):248701--5, December 2012.
\newblock \href {http://dx.doi.org/10.1103/PhysRevLett.109.248701}
  {\path{doi:10.1103/PhysRevLett.109.248701}}.

\bibitem{Cellai2013}
Davide Cellai, Eduardo L\'opez, Jie Zhou, James Gleeson, and Ginestra Bianconi.
\newblock Percolation in multiplex networks with overlap.
\newblock {\em Physical Review E}, 88:052811, Nov 2013.
\newblock URL: \url{http://link.aps.org/doi/10.1103/PhysRevE.88.052811}, \href
  {http://dx.doi.org/10.1103/PhysRevE.88.052811}
  {\path{doi:10.1103/PhysRevE.88.052811}}.

\bibitem{Min2014_robustness_multiplex_interlayer_degree_correlations}
Byungjoon Min, Su~Do Yi, Kyu-Min Lee, and K.-I. Goh.
\newblock Network robustness of multiplex networks with interlayer degree
  correlations.
\newblock {\em Physical Review E}, 89:042811, Apr 2014.
\newblock URL: \url{http://link.aps.org/doi/10.1103/PhysRevE.89.042811}, \href
  {http://dx.doi.org/10.1103/PhysRevE.89.042811}
  {\path{doi:10.1103/PhysRevE.89.042811}}.

\bibitem{Brummitt2012_PRER}
Charles~D Brummitt, Kyu-Min Lee, and K-I Goh.
\newblock {Multiplexity-facilitated cascades in networks}.
\newblock {\em Physical Review E}, 85:045102(R), April 2012.
\newblock \href {http://dx.doi.org/10.1103/PhysRevE.85.045102}
  {\path{doi:10.1103/PhysRevE.85.045102}}.

\bibitem{Yagan2012}
Osman Ya\u{g}an and Virgil Gligor.
\newblock {Analysis of complex contagions in random multiplex networks}.
\newblock {\em Physical Review E}, 86(3):036103, September 2012.
\newblock \href {http://dx.doi.org/10.1103/PhysRevE.86.036103}
  {\path{doi:10.1103/PhysRevE.86.036103}}.

\bibitem{Lee2014}
Kyu-Min Lee, Charles~D Brummitt, and K-I Goh.
\newblock {Threshold cascades with response heterogeneity in multiplex
  networks}.
\newblock {\em Physical Review E}, 90(6), December 2014.
\newblock \href {http://dx.doi.org/10.1103/PhysRevE.90.062816}
  {\path{doi:10.1103/PhysRevE.90.062816}}.

\bibitem{Watts2002}
Duncan~J. Watts.
\newblock A simple model of global cascades on random networks.
\newblock {\em Proceedings of the National Academy of Sciences},
  99(9):5766--5771, 2002.
\newblock \href {http://dx.doi.org/10.1073/pnas.082090499}
  {\path{doi:10.1073/pnas.082090499}}.

\bibitem{Gleeson2007}
JP~Gleeson and DJ~Cahalane.
\newblock {Seed size strongly affects cascades on random networks}.
\newblock {\em Physical Review E}, 75(5):56103, 2007.
\newblock \href {http://dx.doi.org/10.1103/PhysRevE.75.056103}
  {\path{doi:10.1103/PhysRevE.75.056103}}.

\bibitem{Gleeson2008}
James~P. Gleeson.
\newblock {Cascades on correlated and modular random networks}.
\newblock {\em Physical Review E}, 77(4):46117, 2008.
\newblock \href {http://dx.doi.org/10.1103/PhysRevE.77.046117}
  {\path{doi:10.1103/PhysRevE.77.046117}}.

\bibitem{Payne2011}
Joshua Payne, Kameron Harris, and Peter Dodds.
\newblock {Exact solutions for social and biological contagion models on mixed
  directed and undirected, degree-correlated random networks}.
\newblock {\em Physical Review E}, 84(1), July 2011.
\newblock \href {http://dx.doi.org/10.1103/PhysRevE.84.016110}
  {\path{doi:10.1103/PhysRevE.84.016110}}.

\bibitem{Melnik2013}
Sergey Melnik, Jonathan~A Ward, James~P Gleeson, and Mason~A Porter.
\newblock {Multi-stage complex contagions}.
\newblock {\em Chaos: An Interdisciplinary Journal of Nonlinear Science},
  23(1):013124, 2013.
\newblock \href {http://dx.doi.org/10.1063/1.4790836}
  {\path{doi:10.1063/1.4790836}}.

\bibitem{Boss2004}
Michael Boss, Helmut Elsinger, Martin Summer, and Stefan Thurner.
\newblock Network topology of the interbank market.
\newblock {\em Quantitative Finance}, 4(6):677--684, 2004.
\newblock \href {http://dx.doi.org/10.1080/14697680400020325}
  {\path{doi:10.1080/14697680400020325}}.

\bibitem{DeMasi2006}
G.~De~Masi, G.~Iori, and G.~Caldarelli.
\newblock Fitness model for the italian interbank money market.
\newblock {\em Phys. Rev. E}, 74:066112, Dec 2006.
\newblock URL: \url{http://link.aps.org/doi/10.1103/PhysRevE.74.066112}, \href
  {http://dx.doi.org/10.1103/PhysRevE.74.066112}
  {\path{doi:10.1103/PhysRevE.74.066112}}.

\bibitem{Cont2013}
Rama Cont, Amal Moussa, and Edson~B Santos.
\newblock Network structure and systemic risk in banking systems.
\newblock In Jean-Pierre Fouque and Joseph~A. Langsam, editors, {\em Handbook
  on Systemic Risk}. Cambridge University Press, New York, 2013.

\bibitem{Rogers2013}
L~C~G Rogers and L~A~M Veraart.
\newblock {Failure and Rescue in an Interbank Network}.
\newblock {\em Management Science}, 59(4):882--898, April 2013.
\newblock \href {http://dx.doi.org/10.1287/mnsc.1120.1569}
  {\path{doi:10.1287/mnsc.1120.1569}}.

\bibitem{Elsinger2009}
Helmut Elsinger.
\newblock {Financial Networks, Cross Holdings, and Limited Liability }.
\newblock {\em Oesterreichische Nationalbank Working Paper 156}, May 2009.

\bibitem{Gourieroux2013}
C~Gourieroux, J~C Heam, and A~Monfort.
\newblock {Liquidation equilibrium with seniority and hidden CDO}.
\newblock {\em Journal of Banking and Finance}, 37(12):5261--5274, December
  2013.
\newblock \href {http://dx.doi.org/10.1016/j.jbankfin.2013.04.016}
  {\path{doi:10.1016/j.jbankfin.2013.04.016}}.

\bibitem{Eisenberg2001}
L.~Eisenberg and T.~Noe.
\newblock {Systemic risk in financial systems}.
\newblock {\em Management Science}, 47(2):236--249, 2001.
\newblock \href {http://dx.doi.org/10.1287/mnsc.47.2.236.9835}
  {\path{doi:10.1287/mnsc.47.2.236.9835}}.

\bibitem{Centola2007}
D~Centola, V~Egu{\'\i}luz, and M~W Macy.
\newblock {Cascade dynamics of complex propagation}.
\newblock {\em Physica A: Statistical Mechanics and its Applications},
  374(1):449--456, January 2007.
\newblock \href {http://dx.doi.org/doi:10.1016/j.physa.2006.06.018}
  {\path{doi:doi:10.1016/j.physa.2006.06.018}}.

\bibitem{Payne2009}
Joshua Payne, Peter Dodds, and Margaret Eppstein.
\newblock {Information cascades on degree-correlated random networks}.
\newblock {\em Physical Review E}, 80(2), August 2009.
\newblock \href {http://dx.doi.org/10.1103/PhysRevE.80.026125}
  {\path{doi:10.1103/PhysRevE.80.026125}}.

\bibitem{Liu2012}
R~R Liu, W~X Wang, Y~C Lai, and B~H Wang.
\newblock {Cascading dynamics on random networks: Crossover in phase
  transition}.
\newblock {\em Physical Review E}, (85):026110, 2012.
\newblock \href {http://dx.doi.org/10.1103/PhysRevE.85.026110}
  {\path{doi:10.1103/PhysRevE.85.026110}}.

\bibitem{Granovetter1978}
Mark Granovetter.
\newblock {Threshold models of collective behavior}.
\newblock {\em American Journal of Sociology}, 83(6):1420--1443, 1978.
\newblock URL: \url{http://www.jstor.org/stable/2778111}.

\bibitem{Bianconi2013}
G.~Bianconi.
\newblock {Statistical Mechanics of Multiplex networks: Entropy and Overlap}.
\newblock {\em Physical Review E}, 87:062806, 2013.
\newblock \href {http://dx.doi.org/10.1103/PhysRevE.87.062806}
  {\path{doi:10.1103/PhysRevE.87.062806}}.

\bibitem{BankLeverageData}
Data on leverage ratios of banks was accessed at
  \url{http://www.bankregdata.com/allHMmet.asp?met=LEV} on January 22, 2015.
  The source of the data is the Federal Financial Institutions Examination
  Council (FFIEC), available at \url{https://cdr.ffiec.gov/public/}.

\bibitem{Goh2001}
K-I Goh, B~Kahng, and D~Kim.
\newblock {Universal Behavior of Load Distribution in Scale-Free Networks}.
\newblock {\em Physical Review Letters}, 87(27):278701, December 2001.
\newblock \href {http://dx.doi.org/10.1103/PhysRevLett.87.278701}
  {\path{doi:10.1103/PhysRevLett.87.278701}}.

\bibitem{Catanzaro2005_EPJB}
M~Catanzaro and R~Pastor-Satorras.
\newblock {Analytic solution of a static scale-free network model}.
\newblock {\em The European Physical Journal B}, 44(2):241--248, April 2005.
\newblock \href {http://dx.doi.org/10.1140/epjb/e2005-00120-9}
  {\path{doi:10.1140/epjb/e2005-00120-9}}.

\bibitem{footnoteOnStaticModel}
\new{The original static model introduces correlations in the degrees of
  neighbors~\cite{Catanzaro2005_EPJB}. To avoid such correlations, we constrain
  degrees to be less than $\sqrt{N}$ (where $N$ is the number of
  banks)~\cite{Catanzaro2005_PRE}, and we use the configuration model with in-
  and out-degrees drawn from the analytical solution of the degree distribution
  in the static model~\cite[Eq.~(35)]{Catanzaro2005_EPJB}.}

\bibitem{Beale2011}
Nicholas Beale, David~G. Rand, Heather Battey, Karen Croxson, Robert~M. May,
  and Martin~A. Nowak.
\newblock Individual versus systemic risk and the regulator's dilemma.
\newblock {\em Proceedings of the National Academy of Sciences},
  108(31):12647--12652, 2011.
\newblock \href {http://dx.doi.org/10.1073/pnas.1105882108}
  {\path{doi:10.1073/pnas.1105882108}}.

\bibitem{Brummitt2014}
Charles~D Brummitt, Rajiv Sethi, and Duncan~J Watts.
\newblock {Inside Money, Procyclical Leverage, and Banking Catastrophes}.
\newblock {\em PloS one}, 9(8):e104219, August 2014.
\newblock \href {http://dx.doi.org/10.1371/journal.pone.0104219}
  {\path{doi:10.1371/journal.pone.0104219}}.

\bibitem{Caccioli2015}
Fabio Caccioli, J.~Doyne Farmer, Nick Foti, and Daniel Rockmore.
\newblock Overlapping portfolios, contagion, and financial stability.
\newblock {\em Journal of Economic Dynamics and Control}, 51(0):50--63, 2015.
\newblock URL:
  \url{http://www.sciencedirect.com/science/article/pii/S0165188914002632},
  \href {http://dx.doi.org/10.1016/j.jedc.2014.09.041}
  {\path{doi:10.1016/j.jedc.2014.09.041}}.

\bibitem{Catanzaro2005_PRE}
Michele Catanzaro, Mari{\'a}n Bogu{\~n}{\'a}, and Romualdo Pastor-Satorras.
\newblock {Generation of uncorrelated random scale-free networks}.
\newblock {\em Physical Review E}, 71(2):027103, February 2005.
\newblock \href {http://dx.doi.org/10.1103/PhysRevE.71.027103}
  {\path{doi:10.1103/PhysRevE.71.027103}}.

\end{thebibliography}

\end{document}